\documentclass[aip,apl,10pt,reprint]{revtex4-2}
\usepackage{natbib}
\usepackage{hhline}
 
\usepackage{amsfonts}
\usepackage{amsmath}
\usepackage{siunitx}
\usepackage{hyperref}
\usepackage{float}
\usepackage[inline]{enumitem}
\usepackage{tikz,pgfplots}
\usepackage[eulergreek]{sansmath}
\usepackage{upgreek}
\DeclareSymbolFont{sfletters}{OML}{cmbrm}{m}{it}
\DeclareMathSymbol{\sPhi}{\mathord}{sfletters}{"08}

\begin{document}

\title{Purity benchmarking study of error coherence in a single Xmon qubit}

\author{Auda Zhu}
\thanks{The author previously published under the name: Cheng Zhu.}
\affiliation{Institute for Quantum Computing, University of Waterloo, 200 University Avenue West, Waterloo, Ontario N2L 3G1, Canada}
\affiliation{Department of Physics and Astronomy, University of Waterloo, 200 University Avenue West, Waterloo, Ontario N2L 3G1, Canada}

\author{J\'er\'emy H. B\'ejanin}
\affiliation{Institute for Quantum Computing, University of Waterloo, 200 University Avenue West, Waterloo, Ontario N2L 3G1, Canada}
\affiliation{Department of Physics and Astronomy, University of Waterloo, 200 University Avenue West, Waterloo, Ontario N2L 3G1, Canada}

\author{Xicheng Xu}
\affiliation{Institute for Quantum Computing, University of Waterloo, 200 University Avenue West, Waterloo, Ontario N2L 3G1, Canada}
\affiliation{Department of Physics and Astronomy, University of Waterloo, 200 University Avenue West, Waterloo, Ontario N2L 3G1, Canada}

\author{Matteo Mariantoni}
\affiliation{Institute for Quantum Computing, University of Waterloo, 200 University Avenue West, Waterloo, Ontario N2L 3G1, Canada}
\affiliation{Department of Physics and Astronomy, University of Waterloo, 200 University Avenue West, Waterloo, Ontario N2L 3G1, Canada}

\date{\today}

\begin{abstract}
In this study, we employ purity benchmarking (PB) to explore the dynamics of gate noise in a superconducting qubit system. Over 1110 hours of observations on an Xmon qubit, we simultaneously measure the coherence noise budget across two different operational frequencies. We find that incoherent errors, which predominate in overall error rates, exhibit minimal frequency dependence, suggesting they are primarily due to wide-band, diffusive incoherent error sources. In contrast, coherent errors, although less prevalent, show significant sensitivity to operational frequency variations and telegraphic noise. We speculate that this sensitivity is due to interactions with a single strongly coupled environmental defect—--modeled as a two-level system—--which influences qubit control parameters and causes coherent calibration errors. Our results also demonstrate that PB offers improved sensitivity, capturing additional dynamics that conventional relaxation time measurements cannot detect, thus presenting a more comprehensive method for capturing dynamic interactions within quantum systems. The intricate nature of these coherence dynamics underscores the need for further research.
\end{abstract}
\maketitle 

\section{Introduction}

As quantum computing rapidly advances, superconducting qubits emerge as leading candidates for constructing quantum computers\cite{PRXQ_superconducting_circuit}, alongside other contenders such as trapped ions and neutral atoms\cite{PRXQ_trappedions_neutalatoms}. Despite their promise, these systems confront significant challenges, notably in controlling quantum operations within the acceptable error thresholds necessary for practical applications\cite{nature_threshold}. Given these challenges, there is a pressing need for more scalable and practical methods to assess and study the quantum gate errors effectively.

Unlike standard process tomography, which has noticeable inefficiencies\cite{QPT,QPT_2,QPT_3}, randomized benchmarking (RB) offers a more scalable and practical method to assess the gate error by estimating the average gate infidelity $\epsilon$ of gate operations\cite{emerson_scalable_2005,magesan_scalable_2011}. This measure of infidelity is critical as it can be directly used to bound the worst-case error rate, often expressed by the diamond norm\cite{diamond_norm,diamond_norm2} in the context of fault-tolerant computing. The Wallman-Flammia inequalities provide a mathematical framework to relate this norm to average gate infidelity, demonstrating the efficiency and relevance of RB in addressing the challenges of error assessments\cite{WF_bounds}:
\begin{equation}
\frac{D+1}{D}\epsilon\leq\frac{1}{2}||\mathbb{I}-\mathcal{E}||_\diamond\leq\sqrt{D(D+1)\epsilon},
\end{equation}
where $||\mathbb{I}-\mathcal{E}||_\diamond$ is the diamond norm distance of the pure error process $\mathcal{E}$ to the identity, and $D$ is the dimension of the system.

Traditional studies of qubit fluctuation\cite{interactingDefects,T1_1,carroll2022dynamics} have primarily focused on how error process influences phenomena like dephasing ($T_\sPhi$) and energy relaxation ($T_1$), which are typically categorized as incoherent processes. However, recent research\cite{tersoff2021modeling,McRae_review} has begun to explore how these processes might also influence coherent error mechanisms. While most experimental studies still rely on $T_1$, $T_\sPhi$, and qubit frequency $f_{01}$ measurements, a novel technique derived from RB---known as purity benchmarking\cite{wallman_estimating_2015,feng_estimating_2016} (PB)---provides a means to directly monitor the coherence of qubit gate error.

Building on our previous study where we analyzed total gate error ($\epsilon$) fluctuations by performing RB over extended periods, including sessions up to 40 hours\cite{demuxyz}, we now aim to deepen our exploration into the noise budget of total errors. In this study, we conduct PB experiments over time, with a specific focus on the coherence of qubit errors.

Evidence shows that the most significant sources of qubit error and fluctuation can be attributed to interactions with defects in the surrounding amorphous material, which can be modeled as two-level systems (TLSs)\cite{8, 9_2, neeley2008process,burnett2014evidence,GTM,TLSReview}. From our previous research\cite{interactingDefects, LERes}, we have identified two distinct types of TLS interactions: one involves wide-band ensembles of weakly coupled TLSs, leading to diffusive error processes, and the other involves individually strongly coupled TLSs, typically linked to telegraphic error processes.

In this study, we conduct long-time continuous PB measurements simultaneously at two operational frequencies on a single Xmon qubit—one significantly closer to the center frequency of a strongly coupled TLS than the other, utilizing the unique frequency tunability of the Xmon qubit. This approach allows us to differentiate the effects of individual TLS coupling strengths. The experiments extend over an period of 1110 hours, enabling long-term studies on both coherent and incoherent error fluctuations.

In our findings, incoherent errors predominate in the noise budget, indicating that gate operations are primarily vulnerable to incoherent processes. Moreover, the minimal frequency dependency observed in the incoherent error rate aligns with existing studies, suggesting that qubit coherence is predominantly compromised by interactions with wide-band, weakly coupled TLS ensembles.

Interestingly, we find that coherent errors are profoundly influenced by differences in operational frequency. Notably, despite their smaller magnitude, coherent errors account for almost all observed frequency differences. This observation supports the notion that individual strong TLS coupling acts not only as an incoherent error source, as traditionally documented with effects on $T_1$, but also significantly influences coherent errors. We speculate that this dual impact involves individual strongly coupled TLSs switching critical qubit operation parameters, such as frequencies and anharmonicity, leading to calibration issues and significant coherent errors.

Moreover, our results highlight that coherent errors are particularly sensitive to telegraphic noise. Additionally, coherent errors reveal intriguing telegraphic dynamic that traditional $T_1$ measurements fail to detect or explain, underscoring the advantage of PB measurements.

In the following sections, we briefly review the fundamentals of error coherence and PB (Section \ref{sec:theory}), then we introduce our modified experimental protocol (Section \ref{sec:methods}) used to continuously monitor fluctuations in total, coherent, and incoherent errors over time. The results are presented in Section \ref{sec:results} and further analyzed in Section \ref{sec:discussions}, highlighting the unexpected frequency dependencies related to single strongly coupled TLS and the susceptibility of coherent errors to telegraphic noise, which are often obscured by simultaneous $T_1$ measurements. Improvements to the PB protocol, which enhance the accuracy of the measurements, are detailed in the Appendix. Additional experiments that support these findings are included in the supplementary materials.

\section{Theory}
\label{sec:theory}

Over time, multiple metrics for quantum error have been developed to serve different research interests\cite{wallman_estimating_2015,diamond_norm}.  Among these, the concept of average gate fidelity, denoted as $F_{\text{avg}}$ (or $\overline{F}$), has gained prominence, especially with the introduction of RB\cite{emerson_scalable_2005}. This metric, which quantifies the average performance of quantum gates, is determined by calculating the Haar-average of the gate fidelity $F$ over all pure input states, utilizing the Fubini-Study measure across the complete unitary group.  This approach was refined by E. Megason \textit{et al.}, who adapted it to focus on the Clifford groups to develop generalized RB, enhancing its experimental utility and improving RB to a scalable and experimentally practical protocol\cite{magesan_scalable_2011}.

RB operates by implementing sequences of gates randomly drawn from the Clifford group, followed by a gate that reverses the sequence's cumulative effect, aiming to restore the system to its initial state. The protocol quantifies average gate fidelity by measuring the decay in survival probability as the sequence length increases. A principal finding of RB theory is that this decay process can be modeled as a linear combination of depolarizing channels under minor assumptions. By determining the decay rates, RB offers a much efficient method for quantitatively assessing overall gate error through the metric of average gate fidelity. Compared to standard tomography, RB is generally regarded as free of state preparation and measurement (SPAM) errors.

Specifically, for a single qubit, under the assumption that the errors are relatively small and gate-independent, the error process can be simplified to a single depolarizing channel. Consider a single Xmon qubit initialized in its energy ground state $|g\rangle$. After applying $m+1$ gates---$m$ randomly drawn Clifford gates followed by an inverse gate $g_{inv}$---the survival probability, $\overline{P}_g(m+1)$, can be simply described by an exponential decay:

\begin{equation}
\label{Eq:PgExpFit}
\overline{P}_g(m+1) = Ap^m + B,
\end{equation}
where $A$, $B$, and $p$ are the fitting parameters. This model enables the direct calculation of average gate fidelity, $F_{\text{avg}}$, and the average error rate under the same metric, $\epsilon$, as follows:

\begin{equation}
F_{\text{avg}} = \frac{1}{2}p + \frac{1}{2},
\end{equation}
\begin{equation}
\epsilon = 1 - F_{\text{avg}} = \frac{1}{2} - \frac{1}{2}p.
\end{equation}

\subsection{Coherence of Error Processes}
Although the error rate, $\epsilon$, serves as a common index for assessing the overall performance of gate operations, it falls short in shedding light on the specific errors components contributing to the overall error rate. To gain deeper insights, it is beneficial to further categorize errors based on their coherence properties, which essentially reflects the extent to which an error process maintains the coherence of the quantum state.

A coherent quantum operation can usually be represented by a unitary process, making the process's unitarity a critical measure of its coherence. The underlying principle is that unitarity, as a measure, should remain invariant under unitary transformations and reaches its maximum value when the process itself is completely unitary, diminishing otherwise.

To quantify unitarity, one approach involves evaluating the purity decay of a pure state subjected to a quantum operation, given that purity is preserved under unitary operations but diminishes otherwise. A popular and robust definition, as proposed by Wallman \textit{et al.}\cite{wallman_estimating_2015}, calculates unitarity based on the average purity of output states, adjusted by subtracting identity components:
\begin{equation}
    u(\mathcal{E})=\frac{D}{D-1}\int d\psi\; \text{Tr}\Bigl(
    \mathcal{E}'\bigl(\psi\bigr)^\dagger\mathcal{E}'\bigl(\psi\bigr)
    \Bigr),
\end{equation}
where $\mathcal{E}'\bigl(A\bigr) = \mathcal{E}\bigl(A\bigr) - \text{Tr}\mathcal{E}\bigl(A\bigr)/\sqrt{D}\times\mathbb{I}$ applies to all trace-less Kraus operators $A$ for the error process $\mathcal{E}$, integrating over all pure states. $D$ represents the dimension of corresponding Hilbert space. This unitarity measure $u(\mathcal{E})$ achieves its maximal value of $1$ if and only if the process is purely unitary. This metric also establishes an upper limit on the portion of total errors that is theoretically correctable by unitary operations, as quantified by average gate fidelity:
\begin{equation}
\epsilon_\text{coh}(\mathcal{E}) \leq \epsilon(\mathcal{E}) - \frac{D-1}{D}(1-\sqrt{u(\mathcal{E})}).
\end{equation}
By taking the upper bound on this limit, this framework allows for the partition of the total error rate, $\epsilon$, into its incoherent and coherent components, with the incoherent error rate defined as:
\begin{equation}
\epsilon_{\text{inc}}(\mathcal{E}) = \frac{D-1}{D}(1-\sqrt{u(\mathcal{E})}),
\end{equation}
and define the coherent portion of the error as the remainder:
\begin{equation}
\epsilon_{\text{coh}} = \epsilon - \epsilon_{\text{inc}}.
\end{equation}
These parameters, $\epsilon_{\text{coh}}$ and $\epsilon_{\text{inc}}$, can be experimentally measured using an adaptation of the RB method, known as purity (or unitarity) benchmarking\cite{wallman_estimating_2015,feng_estimating_2016}.

\subsection{Purity Benchmarking}

PB offers a technique to estimate error coherence by measuring the unitarity decay of a quantum process, which is indicated by changes in state purity. This method extends the standard RB protocol to specifically assess gate operation unitarity, tracking how the purity of a state evolves through progressively longer sequences consisting of random Clifford gates.

Building upon RB, PB aims to quantify the purity of the quantum state. For a single Xmon qubit, operations are usually conducted in the energy basis, allowing the qubit state to be represented as a vector $\Vec{\alpha}$ on the Bloch sphere:
\begin{equation}
\rho = \frac{1}{2}(\mathbb{I}+\Vec{\alpha}\cdot\Vec{\sigma}),
\end{equation}
where $\Vec{\sigma}$ denotes the unit Pauli vectors. The state's purity is determined by the squared magnitude of $\Vec{\alpha}$:
\begin{equation}
\text{Purity}(\rho)=\text{Tr}(\rho^2) = \frac{1}{2}(1+|\alpha|^2).
\end{equation}
In defining unitarity, we define a purity metric, $\mathcal{P}$, as the purity subtracted by an identity:
\begin{equation}
    \mathcal{P}(\rho) = |\alpha|^2=\langle\sigma_x\rangle^2+\langle\sigma_y\rangle^2+\langle\sigma_z\rangle^2.
    \label{defP}
\end{equation}
As justified by Wallman \textit{et al.}\cite{wallman_estimating_2015}, the average $\mathcal{P}$ decays exponentially with respect to the number of gates $m$ applied, at a rate defined by the gate unitarity $u$:
\begin{equation}
    \overline{\mathcal{P}}(m) = Au^m+B.
    \label{pnu}
\end{equation}
where $A,B$ are fitting parameters. With unitarity $u$ reflecting the coherence of the error, the incoherent and coherent errors for a single qubit are then derived respectively:
\begin{equation}
    \epsilon_{\text{inc}}(\mathcal{E}) = \frac{1}{2}(1-\sqrt{u(\mathcal{E})}),
\end{equation}
\begin{equation}
    \epsilon_{\text{coh}}(\mathcal{E}) = \epsilon_{\text{total}}(\mathcal{E}) - \epsilon_{\text{inc}}(\mathcal{E}).
    \label{errorProp}
\end{equation}

To experimentally measure the rescaled purity $\mathcal{P}$, we need to modify the protocol for the RB as detailed by Feng \textit{et al.}\cite{feng_estimating_2016}. In a typical RB experiment on an Xmon qubit, we initialize the qubit into its nearly pure energy ground state, commonly referred to as the \lq\lq$-z$\rq\rq\,state. We then apply the random gate sequence and measure the survival probability of the qubit in the energy basis ($\hat{\sigma}_z$), which is expressed as: 
\begin{equation}
    \overline{P}_g = \Bigl \langle \frac{\mathbb{I}+ \sigma_z}{2} \Bigr \rangle = \frac{1}{2}+\frac{1}{2}\langle \sigma_z\rangle.
\end{equation}
This measurement directly provides $\langle \hat{\sigma}_z\rangle$. To compute $\mathcal{P}$, we also require measurements of $\langle \hat{\sigma}_x\rangle,\langle \hat{\sigma}_y\rangle$. This can be achieved by repeating each random gate sequence from the RB protocol two additional times, each with a $\pi/2$ rotation compiled into the inverse gate sequence to rotate the measurement basis and effectively measure $\langle \hat{\sigma}_x\rangle$ and $\langle \hat{\sigma}_y\rangle$. The use of Clifford gates ensures that these additional rotations can be efficiently compiled using classical computations. $\mathcal{P}$ for a specific gate sequence is then calculated using three measurements: $\langle \hat{\sigma}_x\rangle$, $\langle \hat{\sigma}_y\rangle$, and $\langle \hat{\sigma}_z\rangle$. We average $\mathcal{P}$ over different random sequences of the same length. The averaged $\overline{\mathcal{P}}$ can then be fitted with Eq \ref{pnu} to calculate the unitarity and hence, coherent and incoherent errors. Further improvements to the protocol and additional modifications, aimed at enhancing accuracy by reducing fitting errors, are detailed in the Appendix.

\section{Methods}
\label{sec:methods}

\begin{figure*}
    \centering
    \includegraphics{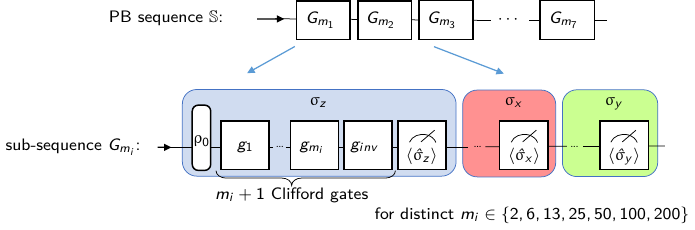}
    \caption[Diagram for PB Sequence.]{Diagram for PB Sequence: each PB sequence $\mathbb{S}$ is composed of $|\mathcal{M}|$ sub-sequences $G$, with each sub-sequence $G$ matched with a distinct $m_i \in \mathcal{M}$. For each $G_{m_i}$, one random sequence of length $m_i+1$ is generated and repeated 3 times. Each time we change the measurement basis by the inverse gate and effectively measure one of $\langle \hat{\sigma}_{z}\rangle$, $\langle \hat{\sigma}_{x}\rangle$, and $\langle \hat{\sigma}_{y}\rangle$.
    }
    \label{fig:PBSequence}
\end{figure*}

\begin{figure}
    \centering
    \includegraphics{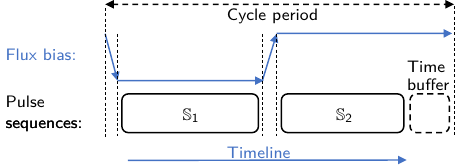}
    \caption[Cycle diagram]{Cycle diagram: Each cycle consists of multiple runs of PB sequences, with each run executed at a different flux bias. This process can be repeated for multiple bias voltages. At the end of each cycle, there is a time buffer to maintain consistent cycle period. By iterating cycles on the qubit, we can monitor errors at different qubit frequencies over time.}
    \label{fig:cycle}
\end{figure}

In response to continuous tracking of time fluctuations of coherent and incoherent errors, we have developed an experimental protocol derived from the existing PB protocol. This adaptation enables detailed, long-term analysis across multiple transition frequencies of a single Xmon qubit.

After initial assessments of the physical qubit setup, we determine a set of appropriate sequence lengths, denoted as $\mathcal{M}=\{2,6,13,25,50,100,200\}$. These gate sequence lengths are chosen to sufficiently capture the exponential decays observed in average gate fidelity and unitarity. It is important to note that the choice of $\mathcal{M}$ is not unique, and the selections result from balancing between the accuracy and efficiency of the measurements. Generally, a larger set or longer gate sequence lengths result in more accurate fittings but require increased measurement times.

\begin{figure*}[ht!]
    \centering
    \includegraphics{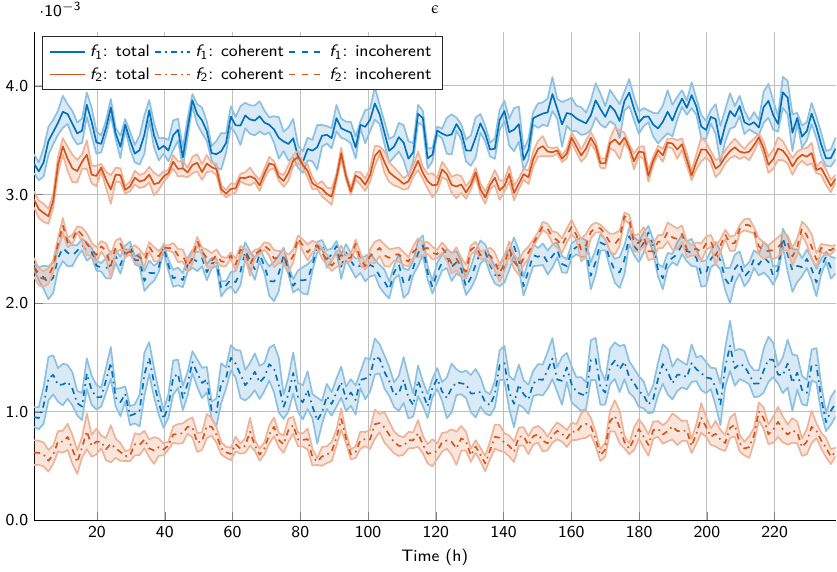}
    \caption{Time series of PB for different errors, shown with shaded error bars. Each data point in this series corresponds to a PB assessment, fitted from data within a moving window covering $n=30$ cycles of measurements. This moving window is designed with a 50\% overlap between consecutive segments. The solid lines depict the total errors $\epsilon$ in two frequencies. The dashed lines represent incoherent components of the errors ($\epsilon_{\text{inc}}$), and the dash-doted lines correspond to the coherent components ($\epsilon_{\text{coh}}$).}
    \label{fig:2140PB}
\end{figure*}

We define a PB sequence $\mathbb{S}$, consisting of $|\mathcal{M}|$ sub-sequences $G$, each paired with a distinct sequence length $m_i \in \mathcal{M}$. For each sub-sequence $G_{m_i}$, a random gate sequence is generated comprising $m_i$ random Clifford gates $g$ followed by an inverse gate $g_{inv}$. This random sequence is then repeated three times. At the end of each repetition, a specific rotation is compiled with the inverse gate to adjust the measurement basis, thereby enabling the measurement of $\langle \hat{\sigma}_{z}\rangle$, $\langle \hat{\sigma}_{x}\rangle$, and $\langle \hat{\sigma}_{y}\rangle$. Figure \ref{fig:PBSequence} illustrates the PB sequence setup, highlighting how each sub-sequence $G_{m_i}$ is constructed and measured. The random Clifford gates are generated by algorithms described in references\cite{JB-thesis,CliffordGen}.

Given the tunability of the Xmon qubit transition frequency via magnetic flux bias, we perform a PB sequence at each flux bias level to estimate errors across multiple frequencies. One execution of PB sequences at each predetermined flux bias point constitutes a cycle, which can be repeated as necessary. To ensure uniform cycle durations, a brief time buffer is introduced at the end of each cycle. A diagram depicting the PB cycle $\mathbb{S}$ is provided in Fig. \ref{fig:cycle} to enhance clarity.

\begin{table*}
    \caption{Summary statistics of the times series data in Fig. \ref{fig:2140PB}. All error metrics are expressed in the scale of $10^{-3}$.}
    \begin{center}
    \begin{ruledtabular}
        \begin{tabular}{lcccccc}
            \raisebox{0mm}[2.5mm][2.0mm]{} & \multicolumn{2}{c}{$\epsilon$}    & \multicolumn{2}{c}{$\epsilon_\text{inc}$} & \multicolumn{2}{c}{$\epsilon_\text{coh}$} \\
            \raisebox{0mm}[3mm][2mm]{} & $f_1$ & $f_2$ & $f_1$ & $f_2$ &$f_1$ & $f_2$ \\
            \hline
            \hline
            \raisebox{0mm}[3mm][1mm]{Mean $\pm$ SD}  & $3.61\pm0.16$ & $3.24\pm0.15$ & $2.36\pm0.11$ & $2.50\pm0.11$ & $1.25\pm0.15$ & $0.74\pm0.10$\\
            \hline
            \raisebox{0mm}[3mm][1mm]{Median} & 3.63 & 3.23 & 2.35 & 2.49 & 1.26 & 0.74\\
            \hline
            \raisebox{0mm}[3.5mm][0mm]{$25^{th}$ to $75^{th}$ percentile} & 0.23 & 0.23 & 0.16 & 0.12 & 0.21 & 0.15\\
        \end{tabular}
    \end{ruledtabular}
    \end{center}
    \label{tab:fitohist}
\end{table*}

After collecting data from multiple cycles, we divide the dataset into windows, each consisting of $n$ PB cycles, and process each window as a separate PB experiment. In every cycle, we measure one random sequence for each sequence length $m_i$. This design enables us to adjust the number $n$ of sequences measured for each length $m_i$ across cycles by varying the window size, thereby balancing measurement time against statistical reliability. Additionally, it enables the use of overlapping windows in our data segmentation, which increases the total number of data points, enhancing the time resolution and improving statistical analysis. The size of each window, $n$, is determined based on preliminary data analysis to optimize the balance between temporal resolution and the statistical significance of the fittings. This method facilitates continuous and reliable measurements over extended periods.

\begin{figure}
    \centering
    \includegraphics{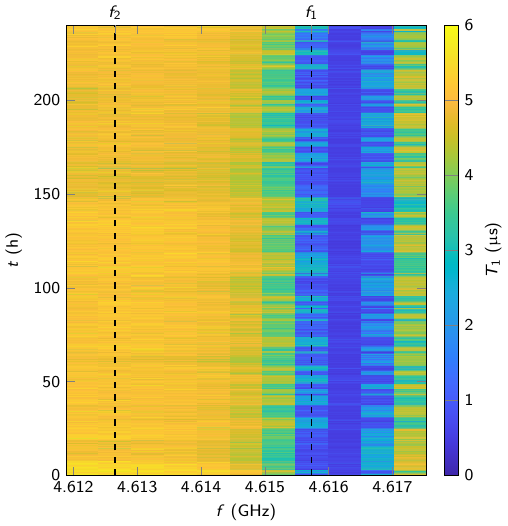}
    \caption{Spectrotemporal chart of $T_1$ vs. $f$ and $t$ for the first experiment. We monitor the qubit $T_1$ in tandem with PB cycles. Two PB frequencies are marked by dashed lines.}
    \label{fig:T1}
\end{figure}

\section{Results}
\label{sec:results}

In this section, we present the results of two representative experiments, each selected to highlight a distinct aspect of our findings: \begin{enumerate*}
    \item Frequency dependency of qubit gate errors due to interactions with single TLS.
    \item Impact of telegraphic events on qubit gate errors.
\end{enumerate*} A concise analysis of these results follows in the subsequent discussion section. 

For the first experiment, Fig. \ref{fig:T1} illustrates the tracking of $T_1$ fluctuations across a range of frequencies over time. Below $4.614\text{ GHz}$, $T_1$ maintains a stable baseline approximately at $5\text{ }\mu\text{s}$. In contrast, near $4.616 \text{ GHz}$, we observe a significant decrease in $T_1$ alongside with spectral-diffusion patterns. These spectrotemporal $T_1$ patterns are clear signs for one strongly coupled quantum TLS, primarily centered around $4.616\text{ GHz}$. Additional swap spectroscopy experiments further corroborate the existence of TLS, aligning well with our previous study\cite{interactingDefects}. 

For the PB measurements, we pick operational frequencies at $f_1 = 4.61530\text{ GHz}$ and $f_2 = 4.61265\text{ GHz}$, with $f_1$ positioned much closer to the strongly coupled TLS to differentiate the TLS coupling strengths. The choice of frequencies $f_1$ and $f_2$ is determined by preliminary measurements. Due to the stochastic nature of the qubit and TLS frequency, we track the operational frequencies over time by applying fixed amount of flux voltages. The applied flux voltages are provided by a battery-powered isolated ultra-clean DC source, the SIM928 (Stanford Research Systems)\cite{SIM}. The input circuits are also heavily isolated, ensuring minimum flux noise to the qubit. The schematics for the entire infrastructure are similar to Fig. 3.5 in reference\cite{JB-thesis}.

\begin{figure}[ht!]
    \centering
    \includegraphics{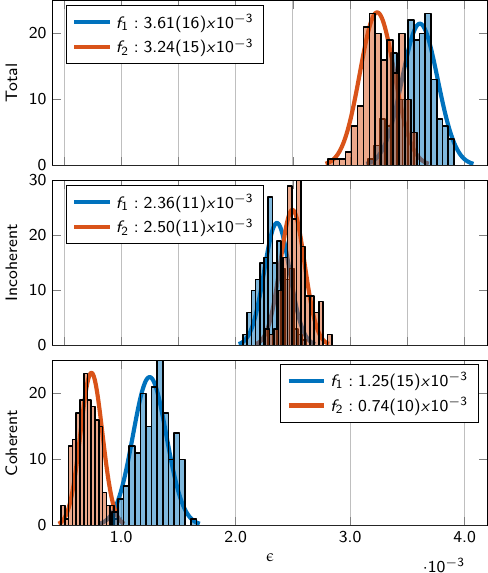}
    \caption{Histograms of time series data in Fig. \ref{fig:2140PB}. The  time series data is grouped into 15 bins and fitted with normal distributions.}
    \label{fig:Hist}
\end{figure}

The data series from PB, with $n = 30$ cycles and 50\% overlapping windows, alongside a corresponding statistical histogram, are displayed in Fig. \ref{fig:2140PB} and Fig. \ref{fig:Hist}, complemented by Table \ref{tab:fitohist}. Error bars indicate 68\% confidence intervals ($1\sigma$). It is noteworthy that the $T_1$ screenings in Fig. \ref{fig:T1} and PB results were obtained in tandem, integrating single $T_1$ measurements within each PB cycle. 

\begin{figure*}
    \centering
    \includegraphics{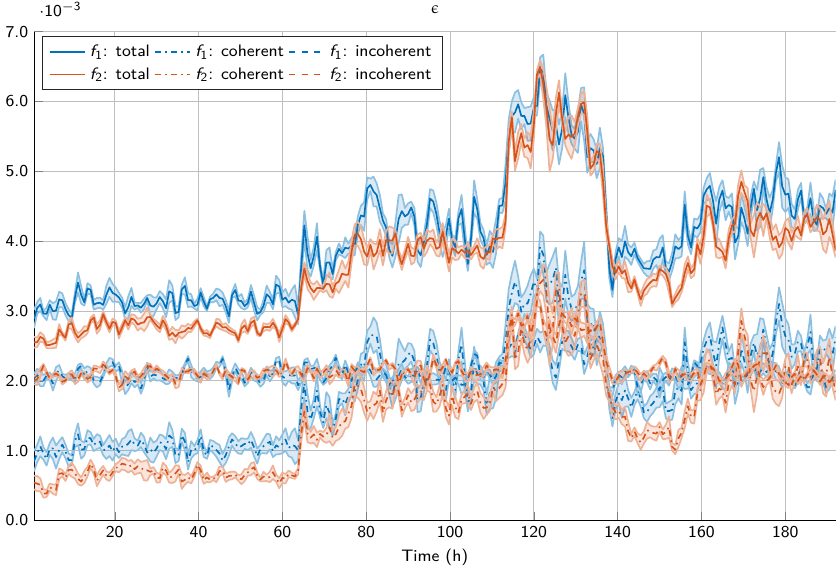}
    \caption{Time series of PB for different errors, shown with shaded error bars. Line styles and color legends are consistent with previous experiment.}
    \label{fig:2136PB}
\end{figure*}

\begin{figure}
    \centering
    \includegraphics{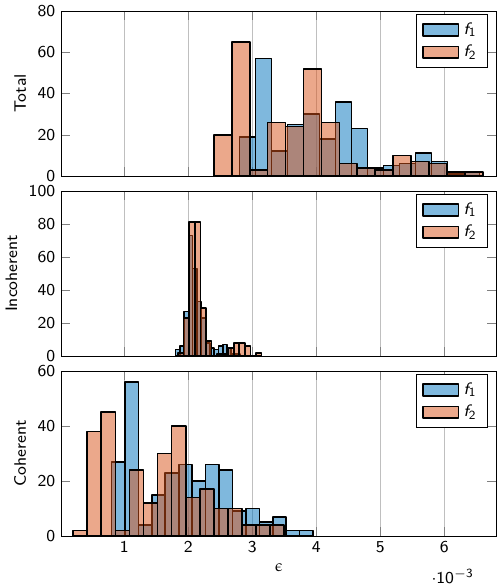}
    \caption{Histograms of time series data for two PB frequencies, $f_1$ and $f_2$.}
    \label{fig:2136hist}
\end{figure}

The second experiment serves as a continuation of the first, following a brief interlude and a thorough recalibration of the qubit control system. This recalibration process includes adjusting the qubit drive frequency, drive pulse amplitude, and the derivative removal by adiabatic gate (DRAG) pulse parameter\cite{DRAG,DRAG2} $\lambda$, while keeping the applied flux voltages for both PB frequencies unchanged. 

\begin{figure}[ht!]
    \centering
    \includegraphics{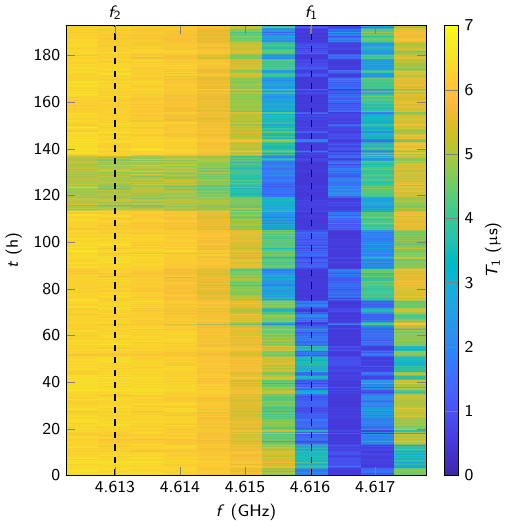}
    \caption{Spectrotemporal chart of $T_1$ vs. $f$ and $t$ in tandem with the second PB experiment.}
    \label{fig:2136T1}
\end{figure}

In Fig. \ref{fig:2136T1}, we present the spectrotemporal $T_1$ chart for the second experiment, employing a similar approach to that of the first experiment. The spectrotemporal $T_1$ patterns still clearly indicate the presence of the strongly coupled TLS. Additionally, a notable, slowly evolving wide-band telegraphic event occurs between approximately 115 and 140 hours. PB time series and complementary histograms are depicted in Fig. \ref{fig:2136PB} and Fig. \ref{fig:2136hist}, respectively. Although both experiments begin with comparable levels of errors, the second experiment exhibits more distinct telegraphic events than the first.  

Additionally, We include results from three additional experiments in the supplementary materials to further corroborating our main analysis. 

\section{Discussions}
\label{sec:discussions}

In this section, we discuss the frequency dependency of qubit coherence in the presence of a strongly coupled TLS, and how telegraphic noise influences these gate errors, building on the results presented previously. While incoherent error represent the major portion in total error, our study emphasize the notable susceptibility of gate coherence to both strongly coupled TLS and telegraphic noises. We propose a speculation for potential future studies. Finally, we highlight coherent errors as novel measurements capable of revealing dynamics that conventional $T_1$ measurements cannot detect, thus opening avenues for future research into the complex interactions between the qubit and its environment.

\subsection{Qualitative Analysis for Single TLS Interaction to Qubit Gate Errors}
Our time series data from both experiments suggest distinct error frequency correlations: total errors intensify with stronger TLS coupling, as operational frequencies align more closely with that of the single strongly coupled TLS. This supports earlier findings that increased proximity to TLS frequency enhances coupling, thereby raising error rates.

From a more quantitative perspective, the majority of the total error is attributed to incoherent errors, which are typically linked to decoherence processes. Among these, TLS couplings are known as major contributors. Interestingly, the incoherent errors across both frequencies appear to be comparable, suggesting that these decoherence processes are less frequency-dependent. Therefore, it is reasonable to assume that incoherent errors are mainly influenced by wide-band weakly coupled TLS ensembles.

On the other hand, while incoherent errors dominate, we find that coherent errors show the most dependency on frequency, suggesting that single strongly coupled TLSs might influence gate operations in a more pronounced coherent manner. Although this strongly coupled TLS exhibits significant incoherent behavior---markedly reducing $T_1$---it is unexpected that the incoherent errors at different frequency proximities to this TLS appear to be similar. This observation suggests a more complex interplay between TLS coupling and qubit gate errors than previously understood.

The hypothesis we propose is that the strongly coupled TLS contributes to the observed increase in coherent errors by altering critical qubit operation parameters, such as transition frequencies and anharmonicity, leading to significant coherent control errors. Additionally, these alterations may be further influenced by the stochastic fluctuations, particularly inherent to these individual strong TLS couplings, which introduce additional noise into the coherent error domain.  These insights lead into the subsequent subsection, which further explores these dynamics.

\subsection{Telegraphic Events on Gate Errors}

In our second experiment, gate operations demonstrate a notable vulnerability to telegraphic noise, with a pronounced impact on coherent errors. This susceptibility can be illustrated by the observed increase in both coherent and total errors between 65 and 110 hours across both frequencies. Unlike conventional $T_1$ measurements, which fail to capture such telegraphic noise, the use of RB/PB techniques proves advantageous.

Additionally, a significant telegraphic jump in incoherent error was observed approximately between 110 to 140 hours. This jump in incoherent error correlates well with the slow wide-band telegraphic decline in $T_1$, further demonstrating the consistency and robustness of the PB techniques in capturing these noises compared to traditional $T_1$ measurements.

The time series data we've gathered, including those presented in the supplementary material, suggest that coherent errors are generally more susceptible to telegraphic noise. If the hypotheses proposed in the previous subsection hold true, they could, to some extent, explain this observed phenomenon: fluctuations in coherent errors are linked to telegraphic error processes by strongly coupled TLS. However, certainty eludes us at this juncture. The underlying physical dynamics at play may be far richer and more intriguing than currently understood, presenting an open question that merits deeper exploration in future studies.

Besides, there is evidence suggesting that the observed increases in all error metrics between 110 and 140 hours, as well as the wide-band reduction in $T_1$, correlate with cosmic activities. According to reports released by the National Aeronautics and Space Administration (NASA), a Coronal Mass Ejection (CME), followed by geomagnetic storms, coincided with the timing of the observed telegraphic error jumps\cite{NASA1,NASA2}. Recent work demonstrates that cosmic rays can strongly impact superconducting qubit systems, leading to faster decoherence and significant increases in correlated errors\cite{martinis2021cosmicrays}. These observations add significant intriguing tangential insights into the experimental results presented in this study.

\section{Conclusion}
In this study, we employed the PB technique to explore the nuanced dynamics of gate noise in quantum systems, with a specific focus on interactions with individually strongly coupled TLS defects. Our experiments have provided insights that, although incoherent errors predominate in overall error rates, they do not demonstrate notable qubit frequency dependence. This phenomenon can be explained by the assumption that the incoherent errors are primarily due to the interactions with the diffusive, weakly coupled TLS ensembles. Interestingly, it is notable that a single strongly coupled TLS, which significantly reduces the qubit’s $T_1$, does not contribute significantly to incoherent errors.

Additionally, we observed that coherent errors, although smaller in magnitude compared to incoherent errors, are markedly sensitive to qubit frequency variations and telegraphic noise. This sensitivity could be attributed to the strongly coupled TLS. Furthermore, we speculate that the dynamics observed in coherent errors may be due to calibration issues caused by the strongly coupled TLS. On the other hand, we also find that coherence measurements reveal richer dynamics which are masked in conventional $T_1$ measurements, demonstrating that PB is a more sensitive and comprehensive technique for understanding qubit-environment interactions.

To further explore this hypothesis, we have simulated gate errors by numerically solving the master equation for a single qubit strongly coupled to a TLS, also taking into account higher level leakages. However, the complexity and scope of these simulations extend beyond the preliminary investigations we have conducted, necessitating far more in-depth studies than can be accommodated within the scope of this article. Thus, we leave these speculations as open questions for future research.

\begin{acknowledgments}
This research was undertaken thanks to funding from the Transformative Quantum Technologies (TQT) program, and was supported by Institute for Quantum Computing (IQC). The authors thank the Quantum-Nano Fabrication and Characterization Facility at the University of Waterloo, where the sample was fabricated.
\end{acknowledgments}
\section*{Author's declarations}

\subsection*{Conflict of interest statement}
The authors have no conflicts to disclose.

\subsection*{Author contributions}
\textbf{Auda Zhu}: Conceptualization (lead); data curation (equal); investigation (lead); methodology (equal); software (equal); formal analysis (lead); visualization (lead); Writing – original draft (lead); writing – review and editing (lead). \textbf{Jérémy H. Benjanin}: Conceptualization (support); data curation (equal); investigation (support); methodology (equal); software (equal); formal analysis (support); visualization (support);. \textbf{Christopher Xu}: Resources - sample fabrication (lead); writing – review and editing (supporting). \textbf{Matteo Mariantoni}: Supervision (lead); validation (lead); funding acquisition (lead); project administration (lead); writing – original draft (supporting); writing – review and editing (supporting)

\section*{Data availability}
The data that support the findings of this study are available from corresponding author upon reasonable request.

\section*{References}

\bibliography{bibliography}

\begin{thebibliography}{37}%
\makeatletter
\providecommand \@ifxundefined [1]{%
 \@ifx{#1\undefined}
}%
\providecommand \@ifnum [1]{%
 \ifnum #1\expandafter \@firstoftwo
 \else \expandafter \@secondoftwo
 \fi
}%
\providecommand \@ifx [1]{%
 \ifx #1\expandafter \@firstoftwo
 \else \expandafter \@secondoftwo
 \fi
}%
\providecommand \natexlab [1]{#1}%
\providecommand \enquote  [1]{``#1''}%
\providecommand \bibnamefont  [1]{#1}%
\providecommand \bibfnamefont [1]{#1}%
\providecommand \citenamefont [1]{#1}%
\providecommand \href@noop [0]{\@secondoftwo}%
\providecommand \href [0]{\begingroup \@sanitize@url \@href}%
\providecommand \@href[1]{\@@startlink{#1}\@@href}%
\providecommand \@@href[1]{\endgroup#1\@@endlink}%
\providecommand \@sanitize@url [0]{\catcode `\\12\catcode `\$12\catcode `\&12\catcode `\#12\catcode `\^12\catcode `\_12\catcode `\%12\relax}%
\providecommand \@@startlink[1]{}%
\providecommand \@@endlink[0]{}%
\providecommand \url  [0]{\begingroup\@sanitize@url \@url }%
\providecommand \@url [1]{\endgroup\@href {#1}{\urlprefix }}%
\providecommand \urlprefix  [0]{URL }%
\providecommand \Eprint [0]{\href }%
\providecommand \doibase [0]{https://doi.org/}%
\providecommand \selectlanguage [0]{\@gobble}%
\providecommand \bibinfo  [0]{\@secondoftwo}%
\providecommand \bibfield  [0]{\@secondoftwo}%
\providecommand \translation [1]{[#1]}%
\providecommand \BibitemOpen [0]{}%
\providecommand \bibitemStop [0]{}%
\providecommand \bibitemNoStop [0]{.\EOS\space}%
\providecommand \EOS [0]{\spacefactor3000\relax}%
\providecommand \BibitemShut  [1]{\csname bibitem#1\endcsname}%
\let\auto@bib@innerbib\@empty
\bibitem [{\citenamefont {Rasmussen}\ \emph {et~al.}(2021)\citenamefont {Rasmussen}, \citenamefont {Christensen}, \citenamefont {Pedersen}, \citenamefont {Kristensen}, \citenamefont {B\ae{}kkegaard}, \citenamefont {Loft},\ and\ \citenamefont {Zinner}}]{PRXQ_superconducting_circuit}%
  \BibitemOpen
  \bibfield  {author} {\bibinfo {author} {\bibfnamefont {S.}~\bibnamefont {Rasmussen}}, \bibinfo {author} {\bibfnamefont {K.}~\bibnamefont {Christensen}}, \bibinfo {author} {\bibfnamefont {S.}~\bibnamefont {Pedersen}}, \bibinfo {author} {\bibfnamefont {L.}~\bibnamefont {Kristensen}}, \bibinfo {author} {\bibfnamefont {T.}~\bibnamefont {B\ae{}kkegaard}}, \bibinfo {author} {\bibfnamefont {N.}~\bibnamefont {Loft}},\ and\ \bibinfo {author} {\bibfnamefont {N.}~\bibnamefont {Zinner}},\ }\bibfield  {title} {\enquote {\bibinfo {title} {Superconducting circuit companion---an introduction with worked examples},}\ }\href {https://doi.org/10.1103/PRXQuantum.2.040204} {\bibfield  {journal} {\bibinfo  {journal} {PRX Quantum}\ }\textbf {\bibinfo {volume} {2}},\ \bibinfo {pages} {040204} (\bibinfo {year} {2021})}\BibitemShut {NoStop}%
\bibitem [{\citenamefont {Alexeev}\ \emph {et~al.}(2021)\citenamefont {Alexeev}, \citenamefont {Bacon}, \citenamefont {Brown}, \citenamefont {Calderbank}, \citenamefont {Carr}, \citenamefont {Chong}, \citenamefont {DeMarco}, \citenamefont {Englund}, \citenamefont {Farhi}, \citenamefont {Fefferman}, \citenamefont {Gorshkov}, \citenamefont {Houck}, \citenamefont {Kim}, \citenamefont {Kimmel}, \citenamefont {Lange}, \citenamefont {Lloyd}, \citenamefont {Lukin}, \citenamefont {Maslov}, \citenamefont {Maunz}, \citenamefont {Monroe}, \citenamefont {Preskill}, \citenamefont {Roetteler}, \citenamefont {Savage},\ and\ \citenamefont {Thompson}}]{PRXQ_trappedions_neutalatoms}%
  \BibitemOpen
  \bibfield  {author} {\bibinfo {author} {\bibfnamefont {Y.}~\bibnamefont {Alexeev}}, \bibinfo {author} {\bibfnamefont {D.}~\bibnamefont {Bacon}}, \bibinfo {author} {\bibfnamefont {K.~R.}\ \bibnamefont {Brown}}, \bibinfo {author} {\bibfnamefont {R.}~\bibnamefont {Calderbank}}, \bibinfo {author} {\bibfnamefont {L.~D.}\ \bibnamefont {Carr}}, \bibinfo {author} {\bibfnamefont {F.~T.}\ \bibnamefont {Chong}}, \bibinfo {author} {\bibfnamefont {B.}~\bibnamefont {DeMarco}}, \bibinfo {author} {\bibfnamefont {D.}~\bibnamefont {Englund}}, \bibinfo {author} {\bibfnamefont {E.}~\bibnamefont {Farhi}}, \bibinfo {author} {\bibfnamefont {B.}~\bibnamefont {Fefferman}}, \bibinfo {author} {\bibfnamefont {A.~V.}\ \bibnamefont {Gorshkov}}, \bibinfo {author} {\bibfnamefont {A.}~\bibnamefont {Houck}}, \bibinfo {author} {\bibfnamefont {J.}~\bibnamefont {Kim}}, \bibinfo {author} {\bibfnamefont {S.}~\bibnamefont {Kimmel}}, \bibinfo {author} {\bibfnamefont {M.}~\bibnamefont {Lange}}, \bibinfo {author} {\bibfnamefont {S.}~\bibnamefont
  {Lloyd}}, \bibinfo {author} {\bibfnamefont {M.~D.}\ \bibnamefont {Lukin}}, \bibinfo {author} {\bibfnamefont {D.}~\bibnamefont {Maslov}}, \bibinfo {author} {\bibfnamefont {P.}~\bibnamefont {Maunz}}, \bibinfo {author} {\bibfnamefont {C.}~\bibnamefont {Monroe}}, \bibinfo {author} {\bibfnamefont {J.}~\bibnamefont {Preskill}}, \bibinfo {author} {\bibfnamefont {M.}~\bibnamefont {Roetteler}}, \bibinfo {author} {\bibfnamefont {M.~J.}\ \bibnamefont {Savage}},\ and\ \bibinfo {author} {\bibfnamefont {J.}~\bibnamefont {Thompson}},\ }\bibfield  {title} {\enquote {\bibinfo {title} {Quantum computer systems for scientific discovery},}\ }\href {https://doi.org/10.1103/PRXQuantum.2.017001} {\bibfield  {journal} {\bibinfo  {journal} {PRX Quantum}\ }\textbf {\bibinfo {volume} {2}},\ \bibinfo {pages} {017001} (\bibinfo {year} {2021})}\BibitemShut {NoStop}%
\bibitem [{\citenamefont {Campbell}(2024)}]{nature_threshold}%
  \BibitemOpen
  \bibfield  {author} {\bibinfo {author} {\bibfnamefont {E.}~\bibnamefont {Campbell}},\ }\bibfield  {title} {\enquote {\bibinfo {title} {A series of fast-paced advances in quantum error correction},}\ }\href {https://doi.org/10.1038/s42254-024-00706-3} {\bibfield  {journal} {\bibinfo  {journal} {Nature Reviews Physics}\ ,\ \bibinfo {pages} {1--2}} (\bibinfo {year} {2024})}\BibitemShut {NoStop}%
\bibitem [{\citenamefont {Chuang}\ and\ \citenamefont {Nielsen}(1997)}]{QPT}%
  \BibitemOpen
  \bibfield  {author} {\bibinfo {author} {\bibfnamefont {I.~L.}\ \bibnamefont {Chuang}}\ and\ \bibinfo {author} {\bibfnamefont {M.~A.}\ \bibnamefont {Nielsen}},\ }\bibfield  {title} {\enquote {\bibinfo {title} {Prescription for experimental determination of the dynamics of a quantum black box},}\ }\href {https://doi.org/10.1080/09500349708231894} {\bibfield  {journal} {\bibinfo  {journal} {Journal of Modern Optics}\ }\textbf {\bibinfo {volume} {44}},\ \bibinfo {pages} {2455--2467} (\bibinfo {year} {1997})}\BibitemShut {NoStop}%
\bibitem [{\citenamefont {Chow}\ \emph {et~al.}(2009)\citenamefont {Chow}, \citenamefont {Gambetta}, \citenamefont {Tornberg}, \citenamefont {Koch}, \citenamefont {Bishop}, \citenamefont {Houck}, \citenamefont {Johnson}, \citenamefont {Frunzio}, \citenamefont {Girvin},\ and\ \citenamefont {Schoelkopf}}]{QPT_2}%
  \BibitemOpen
  \bibfield  {author} {\bibinfo {author} {\bibfnamefont {J.~M.}\ \bibnamefont {Chow}}, \bibinfo {author} {\bibfnamefont {J.~M.}\ \bibnamefont {Gambetta}}, \bibinfo {author} {\bibfnamefont {L.}~\bibnamefont {Tornberg}}, \bibinfo {author} {\bibfnamefont {J.}~\bibnamefont {Koch}}, \bibinfo {author} {\bibfnamefont {L.~S.}\ \bibnamefont {Bishop}}, \bibinfo {author} {\bibfnamefont {A.~A.}\ \bibnamefont {Houck}}, \bibinfo {author} {\bibfnamefont {B.~R.}\ \bibnamefont {Johnson}}, \bibinfo {author} {\bibfnamefont {L.}~\bibnamefont {Frunzio}}, \bibinfo {author} {\bibfnamefont {S.~M.}\ \bibnamefont {Girvin}},\ and\ \bibinfo {author} {\bibfnamefont {R.~J.}\ \bibnamefont {Schoelkopf}},\ }\bibfield  {title} {\enquote {\bibinfo {title} {Randomized benchmarking and process tomography for gate errors in a solid-state qubit},}\ }\href {https://doi.org/10.1103/PhysRevLett.102.090502} {\bibfield  {journal} {\bibinfo  {journal} {Phys. Rev. Lett.}\ }\textbf {\bibinfo {volume} {102}},\ \bibinfo {pages} {090502} (\bibinfo {year}
  {2009})}\BibitemShut {NoStop}%
\bibitem [{\citenamefont {Bialczak}\ \emph {et~al.}(2010)\citenamefont {Bialczak}, \citenamefont {Ansmann}, \citenamefont {Hofheinz}, \citenamefont {Lucero}, \citenamefont {Neeley}, \citenamefont {O’Connell}, \citenamefont {Sank}, \citenamefont {Wang}, \citenamefont {Wenner}, \citenamefont {Steffen} \emph {et~al.}}]{QPT_3}%
  \BibitemOpen
  \bibfield  {author} {\bibinfo {author} {\bibfnamefont {R.~C.}\ \bibnamefont {Bialczak}}, \bibinfo {author} {\bibfnamefont {M.}~\bibnamefont {Ansmann}}, \bibinfo {author} {\bibfnamefont {M.}~\bibnamefont {Hofheinz}}, \bibinfo {author} {\bibfnamefont {E.}~\bibnamefont {Lucero}}, \bibinfo {author} {\bibfnamefont {M.}~\bibnamefont {Neeley}}, \bibinfo {author} {\bibfnamefont {A.~D.}\ \bibnamefont {O’Connell}}, \bibinfo {author} {\bibfnamefont {D.}~\bibnamefont {Sank}}, \bibinfo {author} {\bibfnamefont {H.}~\bibnamefont {Wang}}, \bibinfo {author} {\bibfnamefont {J.}~\bibnamefont {Wenner}}, \bibinfo {author} {\bibfnamefont {M.}~\bibnamefont {Steffen}}, \emph {et~al.},\ }\bibfield  {title} {\enquote {\bibinfo {title} {Quantum process tomography of a universal entangling gate implemented with josephson phase qubits},}\ }\href {https://doi.org/10.1038/nphys1639} {\bibfield  {journal} {\bibinfo  {journal} {Nature Physics}\ }\textbf {\bibinfo {volume} {6}},\ \bibinfo {pages} {409--413} (\bibinfo {year}
  {2010})}\BibitemShut {NoStop}%
\bibitem [{\citenamefont {Emerson}, \citenamefont {Alicki},\ and\ \citenamefont {Życzkowski}(2005)}]{emerson_scalable_2005}%
  \BibitemOpen
  \bibfield  {author} {\bibinfo {author} {\bibfnamefont {J.}~\bibnamefont {Emerson}}, \bibinfo {author} {\bibfnamefont {R.}~\bibnamefont {Alicki}},\ and\ \bibinfo {author} {\bibfnamefont {K.}~\bibnamefont {Życzkowski}},\ }\bibfield  {title} {\enquote {\bibinfo {title} {Scalable noise estimation with random unitary operators},}\ }\href {https://doi.org/10.1088/1464-4266/7/10/021} {\bibfield  {journal} {\bibinfo  {journal} {J. Opt. B: Quantum Semiclass. Opt.}\ }\textbf {\bibinfo {volume} {7}},\ \bibinfo {pages} {S347} (\bibinfo {year} {2005})}\BibitemShut {NoStop}%
\bibitem [{\citenamefont {Magesan}, \citenamefont {Gambetta},\ and\ \citenamefont {Emerson}(2011)}]{magesan_scalable_2011}%
  \BibitemOpen
  \bibfield  {author} {\bibinfo {author} {\bibfnamefont {E.}~\bibnamefont {Magesan}}, \bibinfo {author} {\bibfnamefont {J.~M.}\ \bibnamefont {Gambetta}},\ and\ \bibinfo {author} {\bibfnamefont {J.}~\bibnamefont {Emerson}},\ }\bibfield  {title} {\enquote {\bibinfo {title} {Scalable and robust randomized benchmarking of quantum processes},}\ }\href {https://doi.org/10.1103/PhysRevLett.106.180504} {\bibfield  {journal} {\bibinfo  {journal} {Phys. Rev. Lett.}\ }\textbf {\bibinfo {volume} {106}},\ \bibinfo {pages} {180504} (\bibinfo {year} {2011})},\ \bibinfo {note} {publisher: American Physical Society}\BibitemShut {NoStop}%
\bibitem [{\citenamefont {Watrous}(209)}]{diamond_norm}%
  \BibitemOpen
  \bibfield  {author} {\bibinfo {author} {\bibfnamefont {J.}~\bibnamefont {Watrous}},\ }\bibfield  {title} {\enquote {\bibinfo {title} {Semidefinite programs for completely bounded norms},}\ }\href {https://doi.org/10.4086/toc.2009.v005a011} {\bibfield  {journal} {\bibinfo  {journal} {Theory of Computing}\ }\textbf {\bibinfo {volume} {5}},\ \bibinfo {pages} {217--238} (\bibinfo {year} {209})}\BibitemShut {NoStop}%
\bibitem [{\citenamefont {Johnston}, \citenamefont {Kribs},\ and\ \citenamefont {Paulsen}(2007)}]{diamond_norm2}%
  \BibitemOpen
  \bibfield  {author} {\bibinfo {author} {\bibfnamefont {N.}~\bibnamefont {Johnston}}, \bibinfo {author} {\bibfnamefont {D.~W.}\ \bibnamefont {Kribs}},\ and\ \bibinfo {author} {\bibfnamefont {V.~I.}\ \bibnamefont {Paulsen}},\ }\href@noop {} {\enquote {\bibinfo {title} {Computing stabilized norms for quantum operations via the theory of completely bounded maps},}\ } (\bibinfo {year} {2007}),\ \Eprint {https://arxiv.org/abs/0711.3636} {arXiv:0711.3636 [quant-ph]} \BibitemShut {NoStop}%
\bibitem [{\citenamefont {Wallman}\ and\ \citenamefont {Flammia}(2014)}]{WF_bounds}%
  \BibitemOpen
  \bibfield  {author} {\bibinfo {author} {\bibfnamefont {J.~J.}\ \bibnamefont {Wallman}}\ and\ \bibinfo {author} {\bibfnamefont {S.~T.}\ \bibnamefont {Flammia}},\ }\bibfield  {title} {\enquote {\bibinfo {title} {Randomized benchmarking with confidence},}\ }\href {https://doi.org/10.1088/1367-2630/16/10/103032} {\bibfield  {journal} {\bibinfo  {journal} {New J. Phys.}\ }\textbf {\bibinfo {volume} {16}},\ \bibinfo {pages} {103032} (\bibinfo {year} {2014})},\ \bibinfo {note} {publisher: IOP Publishing}\BibitemShut {NoStop}%
\bibitem [{\citenamefont {B\'ejanin}\ \emph {et~al.}(2021)\citenamefont {B\'ejanin}, \citenamefont {Earnest}, \citenamefont {Sharafeldin},\ and\ \citenamefont {Mariantoni}}]{interactingDefects}%
  \BibitemOpen
  \bibfield  {author} {\bibinfo {author} {\bibfnamefont {J.~H.}\ \bibnamefont {B\'ejanin}}, \bibinfo {author} {\bibfnamefont {C.~T.}\ \bibnamefont {Earnest}}, \bibinfo {author} {\bibfnamefont {A.~S.}\ \bibnamefont {Sharafeldin}},\ and\ \bibinfo {author} {\bibfnamefont {M.}~\bibnamefont {Mariantoni}},\ }\bibfield  {title} {\enquote {\bibinfo {title} {Interacting defects generate stochastic fluctuations in superconducting qubits},}\ }\href {https://doi.org/10.1103/PhysRevB.104.094106} {\bibfield  {journal} {\bibinfo  {journal} {Phys. Rev. B}\ }\textbf {\bibinfo {volume} {104}},\ \bibinfo {pages} {094106} (\bibinfo {year} {2021})}\BibitemShut {NoStop}%
\bibitem [{\citenamefont {Klimov}\ \emph {et~al.}(2018)\citenamefont {Klimov}, \citenamefont {Kelly}, \citenamefont {Chen}, \citenamefont {Neeley}, \citenamefont {Megrant}, \citenamefont {Burkett}, \citenamefont {Barends}, \citenamefont {Arya}, \citenamefont {Chiaro}, \citenamefont {Chen}, \citenamefont {Dunsworth}, \citenamefont {Fowler}, \citenamefont {Foxen}, \citenamefont {Gidney}, \citenamefont {Giustina}, \citenamefont {Graff}, \citenamefont {Huang}, \citenamefont {Jeffrey}, \citenamefont {Lucero}, \citenamefont {Mutus}, \citenamefont {Naaman}, \citenamefont {Neill}, \citenamefont {Quintana}, \citenamefont {Roushan}, \citenamefont {Sank}, \citenamefont {Vainsencher}, \citenamefont {Wenner}, \citenamefont {White}, \citenamefont {Boixo}, \citenamefont {Babbush}, \citenamefont {Smelyanskiy}, \citenamefont {Neven},\ and\ \citenamefont {Martinis}}]{T1_1}%
  \BibitemOpen
  \bibfield  {author} {\bibinfo {author} {\bibfnamefont {P.~V.}\ \bibnamefont {Klimov}}, \bibinfo {author} {\bibfnamefont {J.}~\bibnamefont {Kelly}}, \bibinfo {author} {\bibfnamefont {Z.}~\bibnamefont {Chen}}, \bibinfo {author} {\bibfnamefont {M.}~\bibnamefont {Neeley}}, \bibinfo {author} {\bibfnamefont {A.}~\bibnamefont {Megrant}}, \bibinfo {author} {\bibfnamefont {B.}~\bibnamefont {Burkett}}, \bibinfo {author} {\bibfnamefont {R.}~\bibnamefont {Barends}}, \bibinfo {author} {\bibfnamefont {K.}~\bibnamefont {Arya}}, \bibinfo {author} {\bibfnamefont {B.}~\bibnamefont {Chiaro}}, \bibinfo {author} {\bibfnamefont {Y.}~\bibnamefont {Chen}}, \bibinfo {author} {\bibfnamefont {A.}~\bibnamefont {Dunsworth}}, \bibinfo {author} {\bibfnamefont {A.}~\bibnamefont {Fowler}}, \bibinfo {author} {\bibfnamefont {B.}~\bibnamefont {Foxen}}, \bibinfo {author} {\bibfnamefont {C.}~\bibnamefont {Gidney}}, \bibinfo {author} {\bibfnamefont {M.}~\bibnamefont {Giustina}}, \bibinfo {author} {\bibfnamefont {R.}~\bibnamefont {Graff}},
  \bibinfo {author} {\bibfnamefont {T.}~\bibnamefont {Huang}}, \bibinfo {author} {\bibfnamefont {E.}~\bibnamefont {Jeffrey}}, \bibinfo {author} {\bibfnamefont {E.}~\bibnamefont {Lucero}}, \bibinfo {author} {\bibfnamefont {J.~Y.}\ \bibnamefont {Mutus}}, \bibinfo {author} {\bibfnamefont {O.}~\bibnamefont {Naaman}}, \bibinfo {author} {\bibfnamefont {C.}~\bibnamefont {Neill}}, \bibinfo {author} {\bibfnamefont {C.}~\bibnamefont {Quintana}}, \bibinfo {author} {\bibfnamefont {P.}~\bibnamefont {Roushan}}, \bibinfo {author} {\bibfnamefont {D.}~\bibnamefont {Sank}}, \bibinfo {author} {\bibfnamefont {A.}~\bibnamefont {Vainsencher}}, \bibinfo {author} {\bibfnamefont {J.}~\bibnamefont {Wenner}}, \bibinfo {author} {\bibfnamefont {T.~C.}\ \bibnamefont {White}}, \bibinfo {author} {\bibfnamefont {S.}~\bibnamefont {Boixo}}, \bibinfo {author} {\bibfnamefont {R.}~\bibnamefont {Babbush}}, \bibinfo {author} {\bibfnamefont {V.~N.}\ \bibnamefont {Smelyanskiy}}, \bibinfo {author} {\bibfnamefont {H.}~\bibnamefont {Neven}},\ and\
  \bibinfo {author} {\bibfnamefont {J.~M.}\ \bibnamefont {Martinis}},\ }\bibfield  {title} {\enquote {\bibinfo {title} {Fluctuations of energy-relaxation times in superconducting qubits},}\ }\href {https://doi.org/10.1103/PhysRevLett.121.090502} {\bibfield  {journal} {\bibinfo  {journal} {Phys. Rev. Lett.}\ }\textbf {\bibinfo {volume} {121}},\ \bibinfo {pages} {090502} (\bibinfo {year} {2018})}\BibitemShut {NoStop}%
\bibitem [{\citenamefont {Carroll}\ \emph {et~al.}(2022)\citenamefont {Carroll}, \citenamefont {Rosenblatt}, \citenamefont {Jurcevic}, \citenamefont {Lauer},\ and\ \citenamefont {Kandala}}]{carroll2022dynamics}%
  \BibitemOpen
  \bibfield  {author} {\bibinfo {author} {\bibfnamefont {M.}~\bibnamefont {Carroll}}, \bibinfo {author} {\bibfnamefont {S.}~\bibnamefont {Rosenblatt}}, \bibinfo {author} {\bibfnamefont {P.}~\bibnamefont {Jurcevic}}, \bibinfo {author} {\bibfnamefont {I.}~\bibnamefont {Lauer}},\ and\ \bibinfo {author} {\bibfnamefont {A.}~\bibnamefont {Kandala}},\ }\bibfield  {title} {\enquote {\bibinfo {title} {Dynamics of superconducting qubit relaxation times},}\ }\href@noop {} {\bibfield  {journal} {\bibinfo  {journal} {npj Quantum Information}\ }\textbf {\bibinfo {volume} {8}},\ \bibinfo {pages} {132} (\bibinfo {year} {2022})}\BibitemShut {NoStop}%
\bibitem [{\citenamefont {Tersoff}\ and\ \citenamefont {Hannon}(2021)}]{tersoff2021modeling}%
  \BibitemOpen
  \bibfield  {author} {\bibinfo {author} {\bibfnamefont {J.}~\bibnamefont {Tersoff}}\ and\ \bibinfo {author} {\bibfnamefont {J.}~\bibnamefont {Hannon}},\ }\bibfield  {title} {\enquote {\bibinfo {title} {Modeling of qubit coherence variability due to two-level systems},}\ }in\ \href@noop {} {\emph {\bibinfo {booktitle} {APS March Meeting Abstracts}}},\ Vol.\ \bibinfo {volume} {2021}\ (\bibinfo {year} {2021})\ pp.\ \bibinfo {pages} {J28--003}\BibitemShut {NoStop}%
\bibitem [{\citenamefont {McRae}\ \emph {et~al.}(2021)\citenamefont {McRae}, \citenamefont {Stiehl}, \citenamefont {Wang}, \citenamefont {Lin}, \citenamefont {Caldwell}, \citenamefont {Pappas}, \citenamefont {Mutus},\ and\ \citenamefont {Combes}}]{McRae_review}%
  \BibitemOpen
  \bibfield  {author} {\bibinfo {author} {\bibfnamefont {C.~R.~H.}\ \bibnamefont {McRae}}, \bibinfo {author} {\bibfnamefont {G.~M.}\ \bibnamefont {Stiehl}}, \bibinfo {author} {\bibfnamefont {H.}~\bibnamefont {Wang}}, \bibinfo {author} {\bibfnamefont {S.-X.}\ \bibnamefont {Lin}}, \bibinfo {author} {\bibfnamefont {S.~A.}\ \bibnamefont {Caldwell}}, \bibinfo {author} {\bibfnamefont {D.~P.}\ \bibnamefont {Pappas}}, \bibinfo {author} {\bibfnamefont {J.}~\bibnamefont {Mutus}},\ and\ \bibinfo {author} {\bibfnamefont {J.}~\bibnamefont {Combes}},\ }\bibfield  {title} {\enquote {\bibinfo {title} {{Reproducible coherence characterization of superconducting quantum devices}},}\ }\href {https://doi.org/10.1063/5.0060370} {\bibfield  {journal} {\bibinfo  {journal} {Applied Physics Letters}\ }\textbf {\bibinfo {volume} {119}},\ \bibinfo {pages} {100501} (\bibinfo {year} {2021})}\BibitemShut {NoStop}%
\bibitem [{\citenamefont {Wallman}\ \emph {et~al.}(2015)\citenamefont {Wallman}, \citenamefont {Granade}, \citenamefont {Harper},\ and\ \citenamefont {Flammia}}]{wallman_estimating_2015}%
  \BibitemOpen
  \bibfield  {author} {\bibinfo {author} {\bibfnamefont {J.}~\bibnamefont {Wallman}}, \bibinfo {author} {\bibfnamefont {C.}~\bibnamefont {Granade}}, \bibinfo {author} {\bibfnamefont {R.}~\bibnamefont {Harper}},\ and\ \bibinfo {author} {\bibfnamefont {S.~T.}\ \bibnamefont {Flammia}},\ }\bibfield  {title} {\enquote {\bibinfo {title} {Estimating the coherence of noise},}\ }\href {https://doi.org/10.1088/1367-2630/17/11/113020} {\bibfield  {journal} {\bibinfo  {journal} {New J. Phys.}\ }\textbf {\bibinfo {volume} {17}},\ \bibinfo {pages} {113020} (\bibinfo {year} {2015})}\BibitemShut {NoStop}%
\bibitem [{\citenamefont {Feng}\ \emph {et~al.}(2016)\citenamefont {Feng}, \citenamefont {Wallman}, \citenamefont {Buonacorsi}, \citenamefont {Cho}, \citenamefont {Park}, \citenamefont {Xin}, \citenamefont {Lu}, \citenamefont {Baugh},\ and\ \citenamefont {Laflamme}}]{feng_estimating_2016}%
  \BibitemOpen
  \bibfield  {author} {\bibinfo {author} {\bibfnamefont {G.}~\bibnamefont {Feng}}, \bibinfo {author} {\bibfnamefont {J.~J.}\ \bibnamefont {Wallman}}, \bibinfo {author} {\bibfnamefont {B.}~\bibnamefont {Buonacorsi}}, \bibinfo {author} {\bibfnamefont {F.~H.}\ \bibnamefont {Cho}}, \bibinfo {author} {\bibfnamefont {D.~K.}\ \bibnamefont {Park}}, \bibinfo {author} {\bibfnamefont {T.}~\bibnamefont {Xin}}, \bibinfo {author} {\bibfnamefont {D.}~\bibnamefont {Lu}}, \bibinfo {author} {\bibfnamefont {J.}~\bibnamefont {Baugh}},\ and\ \bibinfo {author} {\bibfnamefont {R.}~\bibnamefont {Laflamme}},\ }\bibfield  {title} {\enquote {\bibinfo {title} {Estimating the coherence of noise in quantum control of a solid-state qubit},}\ }\href {https://doi.org/10.1103/PhysRevLett.117.260501} {\bibfield  {journal} {\bibinfo  {journal} {Phys. Rev. Lett.}\ }\textbf {\bibinfo {volume} {117}},\ \bibinfo {pages} {260501} (\bibinfo {year} {2016})}\BibitemShut {NoStop}%
\bibitem [{\citenamefont {B\'ejanin}, \citenamefont {Earnest},\ and\ \citenamefont {Mariantoni}(2022)}]{demuxyz}%
  \BibitemOpen
  \bibfield  {author} {\bibinfo {author} {\bibfnamefont {J.~H.}\ \bibnamefont {B\'ejanin}}, \bibinfo {author} {\bibfnamefont {C.~T.}\ \bibnamefont {Earnest}},\ and\ \bibinfo {author} {\bibfnamefont {M.}~\bibnamefont {Mariantoni}},\ }\href@noop {} {\enquote {\bibinfo {title} {The quantum socket and demuxyz-based gates with superconducting qubits},}\ } (\bibinfo {year} {2022}),\ \Eprint {https://arxiv.org/abs/2211.00143} {arXiv:2211.00143 [quant-ph]} \BibitemShut {NoStop}%
\bibitem [{\citenamefont {Phillips}(1987)}]{8}%
  \BibitemOpen
  \bibfield  {author} {\bibinfo {author} {\bibfnamefont {W.~A.}\ \bibnamefont {Phillips}},\ }\bibfield  {title} {\enquote {\bibinfo {title} {Two-level states in glasses},}\ }\href {https://doi.org/10.1088/0034-4885/50/12/003} {\bibfield  {journal} {\bibinfo  {journal} {Reports on Progress in Physics}\ }\textbf {\bibinfo {volume} {50}},\ \bibinfo {pages} {1657} (\bibinfo {year} {1987})}\BibitemShut {NoStop}%
\bibitem [{\citenamefont {M{\"u}ller}, \citenamefont {Cole},\ and\ \citenamefont {Lisenfeld}(2019)}]{9_2}%
  \BibitemOpen
  \bibfield  {author} {\bibinfo {author} {\bibfnamefont {C.}~\bibnamefont {M{\"u}ller}}, \bibinfo {author} {\bibfnamefont {J.~H.}\ \bibnamefont {Cole}},\ and\ \bibinfo {author} {\bibfnamefont {J.}~\bibnamefont {Lisenfeld}},\ }\bibfield  {title} {\enquote {\bibinfo {title} {Towards understanding two-level-systems in amorphous solids: insights from quantum circuits},}\ }\href {https://doi.org/10.1088/1361-6633/ab3a7e} {\bibfield  {journal} {\bibinfo  {journal} {Reports on Progress in Physics}\ }\textbf {\bibinfo {volume} {82}},\ \bibinfo {pages} {124501} (\bibinfo {year} {2019})}\BibitemShut {NoStop}%
\bibitem [{\citenamefont {Neeley}\ \emph {et~al.}(2008)\citenamefont {Neeley}, \citenamefont {Ansmann}, \citenamefont {Bialczak}, \citenamefont {Hofheinz}, \citenamefont {Katz}, \citenamefont {Lucero}, \citenamefont {O’connell}, \citenamefont {Wang}, \citenamefont {Cleland},\ and\ \citenamefont {Martinis}}]{neeley2008process}%
  \BibitemOpen
  \bibfield  {author} {\bibinfo {author} {\bibfnamefont {M.}~\bibnamefont {Neeley}}, \bibinfo {author} {\bibfnamefont {M.}~\bibnamefont {Ansmann}}, \bibinfo {author} {\bibfnamefont {R.~C.}\ \bibnamefont {Bialczak}}, \bibinfo {author} {\bibfnamefont {M.}~\bibnamefont {Hofheinz}}, \bibinfo {author} {\bibfnamefont {N.}~\bibnamefont {Katz}}, \bibinfo {author} {\bibfnamefont {E.}~\bibnamefont {Lucero}}, \bibinfo {author} {\bibfnamefont {A.}~\bibnamefont {O’connell}}, \bibinfo {author} {\bibfnamefont {H.}~\bibnamefont {Wang}}, \bibinfo {author} {\bibfnamefont {A.~N.}\ \bibnamefont {Cleland}},\ and\ \bibinfo {author} {\bibfnamefont {J.~M.}\ \bibnamefont {Martinis}},\ }\bibfield  {title} {\enquote {\bibinfo {title} {Process tomography of quantum memory in a josephson-phase qubit coupled to a two-level state},}\ }\href {https://doi.org/10.1038/nphys972} {\bibfield  {journal} {\bibinfo  {journal} {Nature Physics}\ }\textbf {\bibinfo {volume} {4}},\ \bibinfo {pages} {523--526} (\bibinfo {year} {2008})}\BibitemShut
  {NoStop}%
\bibitem [{\citenamefont {Burnett}\ \emph {et~al.}(2014)\citenamefont {Burnett}, \citenamefont {Faoro}, \citenamefont {Wisby}, \citenamefont {Gurtovoi}, \citenamefont {Chernykh}, \citenamefont {Mikhailov}, \citenamefont {Tulin}, \citenamefont {Shaikhaidarov}, \citenamefont {Antonov}, \citenamefont {Meeson} \emph {et~al.}}]{burnett2014evidence}%
  \BibitemOpen
  \bibfield  {author} {\bibinfo {author} {\bibfnamefont {J.}~\bibnamefont {Burnett}}, \bibinfo {author} {\bibfnamefont {L.}~\bibnamefont {Faoro}}, \bibinfo {author} {\bibfnamefont {I.}~\bibnamefont {Wisby}}, \bibinfo {author} {\bibfnamefont {V.}~\bibnamefont {Gurtovoi}}, \bibinfo {author} {\bibfnamefont {A.}~\bibnamefont {Chernykh}}, \bibinfo {author} {\bibfnamefont {G.}~\bibnamefont {Mikhailov}}, \bibinfo {author} {\bibfnamefont {V.}~\bibnamefont {Tulin}}, \bibinfo {author} {\bibfnamefont {R.}~\bibnamefont {Shaikhaidarov}}, \bibinfo {author} {\bibfnamefont {V.}~\bibnamefont {Antonov}}, \bibinfo {author} {\bibfnamefont {P.}~\bibnamefont {Meeson}}, \emph {et~al.},\ }\bibfield  {title} {\enquote {\bibinfo {title} {Evidence for interacting two-level systems from the 1/f noise of a superconducting resonator},}\ }\href {https://doi.org/10.1038/ncomms5119} {\bibfield  {journal} {\bibinfo  {journal} {Nature communications}\ }\textbf {\bibinfo {volume} {5}},\ \bibinfo {pages} {4119} (\bibinfo {year}
  {2014})}\BibitemShut {NoStop}%
\bibitem [{\citenamefont {Faoro}\ and\ \citenamefont {Ioffe}(2015)}]{GTM}%
  \BibitemOpen
  \bibfield  {author} {\bibinfo {author} {\bibfnamefont {L.}~\bibnamefont {Faoro}}\ and\ \bibinfo {author} {\bibfnamefont {L.~B.}\ \bibnamefont {Ioffe}},\ }\bibfield  {title} {\enquote {\bibinfo {title} {Interacting tunneling model for two-level systems in amorphous materials and its predictions for their dephasing and noise in superconducting microresonators},}\ }\href {https://doi.org/10.1103/PhysRevB.91.014201} {\bibfield  {journal} {\bibinfo  {journal} {Phys. Rev. B}\ }\textbf {\bibinfo {volume} {91}},\ \bibinfo {pages} {014201} (\bibinfo {year} {2015})}\BibitemShut {NoStop}%
\bibitem [{\citenamefont {M\"uller}\ \emph {et~al.}(2015)\citenamefont {M\"uller}, \citenamefont {Lisenfeld}, \citenamefont {Shnirman},\ and\ \citenamefont {Poletto}}]{TLSReview}%
  \BibitemOpen
  \bibfield  {author} {\bibinfo {author} {\bibfnamefont {C.}~\bibnamefont {M\"uller}}, \bibinfo {author} {\bibfnamefont {J.}~\bibnamefont {Lisenfeld}}, \bibinfo {author} {\bibfnamefont {A.}~\bibnamefont {Shnirman}},\ and\ \bibinfo {author} {\bibfnamefont {S.}~\bibnamefont {Poletto}},\ }\bibfield  {title} {\enquote {\bibinfo {title} {Interacting two-level defects as sources of fluctuating high-frequency noise in superconducting circuits},}\ }\href {https://doi.org/10.1103/PhysRevB.92.035442} {\bibfield  {journal} {\bibinfo  {journal} {Phys. Rev. B}\ }\textbf {\bibinfo {volume} {92}},\ \bibinfo {pages} {035442} (\bibinfo {year} {2015})}\BibitemShut {NoStop}%
\bibitem [{\citenamefont {B\'ejanin}\ \emph {et~al.}(2022)\citenamefont {B\'ejanin}, \citenamefont {Ayadi}, \citenamefont {Xu}, \citenamefont {Zhu}, \citenamefont {Mohebbi},\ and\ \citenamefont {Mariantoni}}]{LERes}%
  \BibitemOpen
  \bibfield  {author} {\bibinfo {author} {\bibfnamefont {J.~H.}\ \bibnamefont {B\'ejanin}}, \bibinfo {author} {\bibfnamefont {Y.}~\bibnamefont {Ayadi}}, \bibinfo {author} {\bibfnamefont {X.}~\bibnamefont {Xu}}, \bibinfo {author} {\bibfnamefont {C.}~\bibnamefont {Zhu}}, \bibinfo {author} {\bibfnamefont {H.}~\bibnamefont {Mohebbi}},\ and\ \bibinfo {author} {\bibfnamefont {M.}~\bibnamefont {Mariantoni}},\ }\bibfield  {title} {\enquote {\bibinfo {title} {Fluctuation spectroscopy of two-level systems in superconducting resonators},}\ }\href {https://doi.org/10.1103/PhysRevApplied.18.034009} {\bibfield  {journal} {\bibinfo  {journal} {Phys. Rev. Appl.}\ }\textbf {\bibinfo {volume} {18}},\ \bibinfo {pages} {034009} (\bibinfo {year} {2022})}\BibitemShut {NoStop}%
\bibitem [{\citenamefont {B\'ejanin}(2022)}]{JB-thesis}%
  \BibitemOpen
  \bibfield  {author} {\bibinfo {author} {\bibfnamefont {J.~H.}\ \bibnamefont {B\'ejanin}},\ }\emph {\bibinfo {title} {Advances in Superconducting Circuit Quantum Electrodynamics}},\ \href {https://uwspace.uwaterloo.ca/handle/10012/18018} {\bibinfo {type} {{PhD} {Thesis}}},\ \bibinfo  {school} {University of Waterloo} (\bibinfo {year} {2022})\BibitemShut {NoStop}%
\bibitem [{\citenamefont {Koenig}\ and\ \citenamefont {Smolin}(2014)}]{CliffordGen}%
  \BibitemOpen
  \bibfield  {author} {\bibinfo {author} {\bibfnamefont {R.}~\bibnamefont {Koenig}}\ and\ \bibinfo {author} {\bibfnamefont {J.~A.}\ \bibnamefont {Smolin}},\ }\bibfield  {title} {\enquote {\bibinfo {title} {How to efficiently select an arbitrary clifford group element},}\ }\href {https://doi.org/10.1063/1.4903507} {\bibfield  {journal} {\bibinfo  {journal} {Journal of Mathematical Physics}\ }\textbf {\bibinfo {volume} {55}} (\bibinfo {year} {2014}),\ 10.1063/1.4903507}\BibitemShut {NoStop}%
\bibitem [{\citenamefont {{Stanford Research Systems}}(2023)}]{SIM}%
  \BibitemOpen
  \bibfield  {author} {\bibinfo {author} {\bibnamefont {{Stanford Research Systems}}},\ }\href@noop {} {\enquote {\bibinfo {title} {{SIM928 Isolated Voltage Source}},}\ }\bibinfo {howpublished} {\url{https://www.thinksrs.com/products/sim928.html}} (\bibinfo {year} {2023}),\ \bibinfo {note} {accessed: 2023-04-28}\BibitemShut {NoStop}%
\bibitem [{\citenamefont {Motzoi}\ \emph {et~al.}(2009)\citenamefont {Motzoi}, \citenamefont {Gambetta}, \citenamefont {Rebentrost},\ and\ \citenamefont {Wilhelm}}]{DRAG}%
  \BibitemOpen
  \bibfield  {author} {\bibinfo {author} {\bibfnamefont {F.}~\bibnamefont {Motzoi}}, \bibinfo {author} {\bibfnamefont {J.~M.}\ \bibnamefont {Gambetta}}, \bibinfo {author} {\bibfnamefont {P.}~\bibnamefont {Rebentrost}},\ and\ \bibinfo {author} {\bibfnamefont {F.~K.}\ \bibnamefont {Wilhelm}},\ }\bibfield  {title} {\enquote {\bibinfo {title} {Simple pulses for elimination of leakage in weakly nonlinear qubits},}\ }\href {https://doi.org/10.1103/PhysRevLett.103.110501} {\bibfield  {journal} {\bibinfo  {journal} {Phys. Rev. Lett.}\ }\textbf {\bibinfo {volume} {103}},\ \bibinfo {pages} {110501} (\bibinfo {year} {2009})}\BibitemShut {NoStop}%
\bibitem [{\citenamefont {Chen}\ \emph {et~al.}(2016)\citenamefont {Chen}, \citenamefont {Kelly}, \citenamefont {Quintana}, \citenamefont {Barends}, \citenamefont {Campbell}, \citenamefont {Chen}, \citenamefont {Chiaro}, \citenamefont {Dunsworth}, \citenamefont {Fowler}, \citenamefont {Lucero}, \citenamefont {Jeffrey}, \citenamefont {Megrant}, \citenamefont {Mutus}, \citenamefont {Neeley}, \citenamefont {Neill}, \citenamefont {O'Malley}, \citenamefont {Roushan}, \citenamefont {Sank}, \citenamefont {Vainsencher}, \citenamefont {Wenner}, \citenamefont {White}, \citenamefont {Korotkov},\ and\ \citenamefont {Martinis}}]{DRAG2}%
  \BibitemOpen
  \bibfield  {author} {\bibinfo {author} {\bibfnamefont {Z.}~\bibnamefont {Chen}}, \bibinfo {author} {\bibfnamefont {J.}~\bibnamefont {Kelly}}, \bibinfo {author} {\bibfnamefont {C.}~\bibnamefont {Quintana}}, \bibinfo {author} {\bibfnamefont {R.}~\bibnamefont {Barends}}, \bibinfo {author} {\bibfnamefont {B.}~\bibnamefont {Campbell}}, \bibinfo {author} {\bibfnamefont {Y.}~\bibnamefont {Chen}}, \bibinfo {author} {\bibfnamefont {B.}~\bibnamefont {Chiaro}}, \bibinfo {author} {\bibfnamefont {A.}~\bibnamefont {Dunsworth}}, \bibinfo {author} {\bibfnamefont {A.~G.}\ \bibnamefont {Fowler}}, \bibinfo {author} {\bibfnamefont {E.}~\bibnamefont {Lucero}}, \bibinfo {author} {\bibfnamefont {E.}~\bibnamefont {Jeffrey}}, \bibinfo {author} {\bibfnamefont {A.}~\bibnamefont {Megrant}}, \bibinfo {author} {\bibfnamefont {J.}~\bibnamefont {Mutus}}, \bibinfo {author} {\bibfnamefont {M.}~\bibnamefont {Neeley}}, \bibinfo {author} {\bibfnamefont {C.}~\bibnamefont {Neill}}, \bibinfo {author} {\bibfnamefont {P.~J.~J.}\ \bibnamefont
  {O'Malley}}, \bibinfo {author} {\bibfnamefont {P.}~\bibnamefont {Roushan}}, \bibinfo {author} {\bibfnamefont {D.}~\bibnamefont {Sank}}, \bibinfo {author} {\bibfnamefont {A.}~\bibnamefont {Vainsencher}}, \bibinfo {author} {\bibfnamefont {J.}~\bibnamefont {Wenner}}, \bibinfo {author} {\bibfnamefont {T.~C.}\ \bibnamefont {White}}, \bibinfo {author} {\bibfnamefont {A.~N.}\ \bibnamefont {Korotkov}},\ and\ \bibinfo {author} {\bibfnamefont {J.~M.}\ \bibnamefont {Martinis}},\ }\bibfield  {title} {\enquote {\bibinfo {title} {Measuring and suppressing quantum state leakage in a superconducting qubit},}\ }\href {https://doi.org/10.1103/PhysRevLett.116.020501} {\bibfield  {journal} {\bibinfo  {journal} {Phys. Rev. Lett.}\ }\textbf {\bibinfo {volume} {116}},\ \bibinfo {pages} {020501} (\bibinfo {year} {2016})}\BibitemShut {NoStop}%
\bibitem [{\citenamefont {{NASA}}(2023{\natexlab{a}})}]{NASA1}%
  \BibitemOpen
  \bibfield  {author} {\bibinfo {author} {\bibnamefont {{NASA}}},\ }\href@noop {} {\enquote {\bibinfo {title} {Cme event on 2023-03-23},}\ }\bibinfo {howpublished} {\url{https://kauai.ccmc.gsfc.nasa.gov/DONKI/view/CME/24321/1}} (\bibinfo {year} {2023}{\natexlab{a}}),\ \bibinfo {note} {accessed: 2023-04-25}\BibitemShut {NoStop}%
\bibitem [{\citenamefont {{NASA}}(2023{\natexlab{b}})}]{NASA2}%
  \BibitemOpen
  \bibfield  {author} {\bibinfo {author} {\bibnamefont {{NASA}}},\ }\href@noop {} {\enquote {\bibinfo {title} {Cme event on 2023-03-21},}\ }\bibinfo {howpublished} {\url{https://kauai.ccmc.gsfc.nasa.gov/DONKI/view/CME/24313/1}} (\bibinfo {year} {2023}{\natexlab{b}}),\ \bibinfo {note} {accessed: 2023-04-25}\BibitemShut {NoStop}%
\bibitem [{\citenamefont {Martinis}(2021)}]{martinis2021cosmicrays}%
  \BibitemOpen
  \bibfield  {author} {\bibinfo {author} {\bibfnamefont {J.~M.}\ \bibnamefont {Martinis}},\ }\href@noop {} {\enquote {\bibinfo {title} {Saving superconducting quantum processors from qubit decay and correlated errors generated by gamma and cosmic rays},}\ } (\bibinfo {year} {2021}),\ \Eprint {https://arxiv.org/abs/2012.06137} {arXiv:2012.06137 [quant-ph]} \BibitemShut {NoStop}%
\bibitem [{\citenamefont {Harper}\ \emph {et~al.}(2019)\citenamefont {Harper}, \citenamefont {Hincks}, \citenamefont {Ferrie}, \citenamefont {Flammia},\ and\ \citenamefont {Wallman}}]{harper_statistical_2019}%
  \BibitemOpen
  \bibfield  {author} {\bibinfo {author} {\bibfnamefont {R.}~\bibnamefont {Harper}}, \bibinfo {author} {\bibfnamefont {I.}~\bibnamefont {Hincks}}, \bibinfo {author} {\bibfnamefont {C.}~\bibnamefont {Ferrie}}, \bibinfo {author} {\bibfnamefont {S.~T.}\ \bibnamefont {Flammia}},\ and\ \bibinfo {author} {\bibfnamefont {J.~J.}\ \bibnamefont {Wallman}},\ }\bibfield  {title} {\enquote {\bibinfo {title} {Statistical analysis of randomized benchmarking},}\ }\href {https://doi.org/10.1103/PhysRevA.99.052350} {\bibfield  {journal} {\bibinfo  {journal} {Phys. Rev. A}\ }\textbf {\bibinfo {volume} {99}},\ \bibinfo {pages} {052350} (\bibinfo {year} {2019})}\BibitemShut {NoStop}%
\bibitem [{\citenamefont {Muhonen}\ \emph {et~al.}(2015)\citenamefont {Muhonen}, \citenamefont {Laucht}, \citenamefont {Simmons}, \citenamefont {Dehollain}, \citenamefont {Kalra}, \citenamefont {Hudson}, \citenamefont {Freer}, \citenamefont {Itoh}, \citenamefont {Jamieson}, \citenamefont {McCallum}, \citenamefont {Dzurak},\ and\ \citenamefont {Morello}}]{Muhonen_2015}%
  \BibitemOpen
  \bibfield  {author} {\bibinfo {author} {\bibfnamefont {J.~T.}\ \bibnamefont {Muhonen}}, \bibinfo {author} {\bibfnamefont {A.}~\bibnamefont {Laucht}}, \bibinfo {author} {\bibfnamefont {S.}~\bibnamefont {Simmons}}, \bibinfo {author} {\bibfnamefont {J.~P.}\ \bibnamefont {Dehollain}}, \bibinfo {author} {\bibfnamefont {R.}~\bibnamefont {Kalra}}, \bibinfo {author} {\bibfnamefont {F.~E.}\ \bibnamefont {Hudson}}, \bibinfo {author} {\bibfnamefont {S.}~\bibnamefont {Freer}}, \bibinfo {author} {\bibfnamefont {K.~M.}\ \bibnamefont {Itoh}}, \bibinfo {author} {\bibfnamefont {D.~N.}\ \bibnamefont {Jamieson}}, \bibinfo {author} {\bibfnamefont {J.~C.}\ \bibnamefont {McCallum}}, \bibinfo {author} {\bibfnamefont {A.~S.}\ \bibnamefont {Dzurak}},\ and\ \bibinfo {author} {\bibfnamefont {A.}~\bibnamefont {Morello}},\ }\bibfield  {title} {\enquote {\bibinfo {title} {Quantifying the quantum gate fidelity of single-atom spin qubits in silicon by randomized benchmarking},}\ }\href {https://doi.org/10.1088/0953-8984/27/15/154205}
  {\bibfield  {journal} {\bibinfo  {journal} {Journal of Physics: Condensed Matter}\ }\textbf {\bibinfo {volume} {27}},\ \bibinfo {pages} {154205} (\bibinfo {year} {2015})}\BibitemShut {NoStop}%
\bibitem [{\citenamefont {Fogarty}\ \emph {et~al.}(2015)\citenamefont {Fogarty}, \citenamefont {Veldhorst}, \citenamefont {Harper}, \citenamefont {Yang}, \citenamefont {Bartlett}, \citenamefont {Flammia},\ and\ \citenamefont {Dzurak}}]{PhysRevA.92.022326}%
  \BibitemOpen
  \bibfield  {author} {\bibinfo {author} {\bibfnamefont {M.~A.}\ \bibnamefont {Fogarty}}, \bibinfo {author} {\bibfnamefont {M.}~\bibnamefont {Veldhorst}}, \bibinfo {author} {\bibfnamefont {R.}~\bibnamefont {Harper}}, \bibinfo {author} {\bibfnamefont {C.~H.}\ \bibnamefont {Yang}}, \bibinfo {author} {\bibfnamefont {S.~D.}\ \bibnamefont {Bartlett}}, \bibinfo {author} {\bibfnamefont {S.~T.}\ \bibnamefont {Flammia}},\ and\ \bibinfo {author} {\bibfnamefont {A.~S.}\ \bibnamefont {Dzurak}},\ }\bibfield  {title} {\enquote {\bibinfo {title} {Nonexponential fidelity decay in randomized benchmarking with low-frequency noise},}\ }\href {https://doi.org/10.1103/PhysRevA.92.022326} {\bibfield  {journal} {\bibinfo  {journal} {Phys. Rev. A}\ }\textbf {\bibinfo {volume} {92}},\ \bibinfo {pages} {022326} (\bibinfo {year} {2015})}\BibitemShut {NoStop}%
\end{thebibliography}%

\appendix*
\section{Elimination of Offset Terms}
\label{sec:appx}
Engineering advancements and improved noise suppression techniques have significantly reduced the magnitude and fluctuations of quantum errors. However, these advancements also introduce new challenges in capturing the dynamic behavior of these reduced errors over time. The inherent randomness within the PB protocol can induce substantial fitting uncertainty, which may cover the less impacting noise channels and render observed fluctuations seemingly unreliable. In response, research on RB and its variants\cite{harper_statistical_2019,Muhonen_2015,PhysRevA.92.022326} suggests strategies aimed at minimizing fitting errors, such as eliminating offset terms in exponential fittings (e.g., $B$ in Eq \ref{Eq:PgExpFit} and \ref{pnu}).

\begin{figure*}[ht!]
    \centering
    \includegraphics{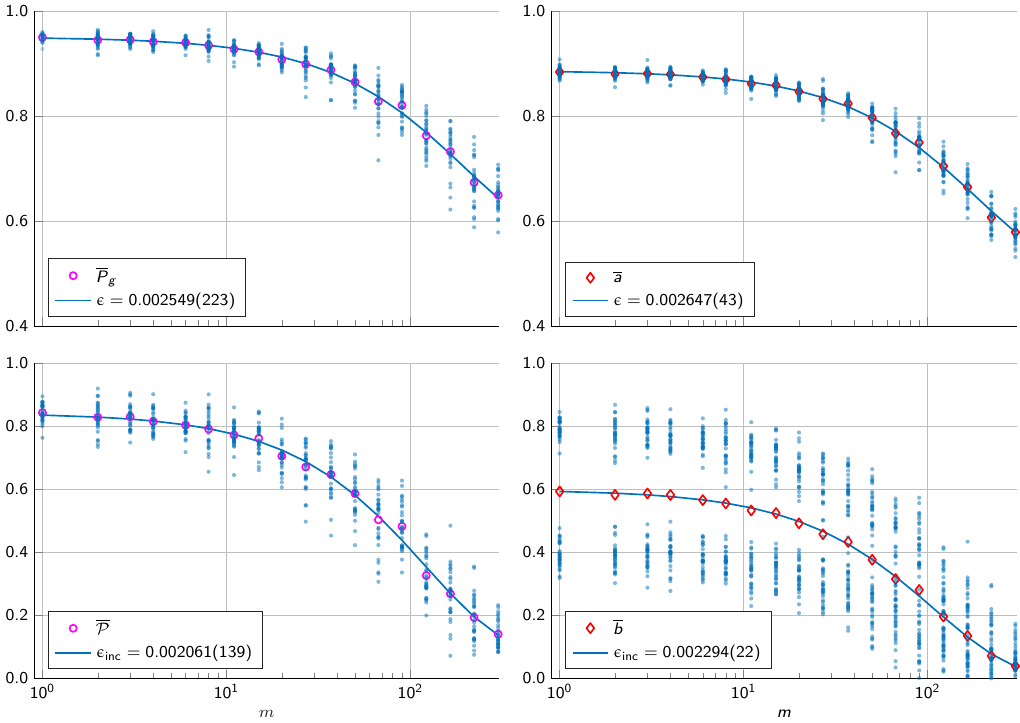}
         \caption{Statistical fittings for the proof-of-concept PB experiment across four different quantities, illustrated in a 2x2 subplot arrangement. Semi-transparent dots in each plot represent individual measurement points for various quantity values, while distinct markers denote the mean values for each quantity at specific sequence lengths ($m$). Solid lines represent fitting curves to these mean values. The left column displays results from PB measurements conducted using standard methods, and the right column shows outcomes from PB measurements after applying the offset elimination technique.}
    \label{fig:appx}
\end{figure*}

\begin{table}[hb!]
    \caption{A proof-of-concept PB experiment demonstrating the improvements of offset eliminations on fitting errors. The fitting errors, quantified by the coefficient of variation (CV), are significantly reduced in all error types.}
    \begin{center}
    \begin{ruledtabular}
        \begin{tabular}{ccc}
            \raisebox{0mm}[3.5mm][3.0mm]{Error type}& With offsets    & Offsets eliminated\\
            \hline
            \hline
            \\[-3.0mm]
            \raisebox{0mm}[3mm][0mm]{$\epsilon$}                       & 8.75 & 1.62 \\
            \hline
            \raisebox{0mm}[3mm][0mm]{$\epsilon_{\text{inc}}$}                  & 6.74 & 0.96\\
            \hline
            \raisebox{0mm}[3mm][0mm]{$\epsilon_{\text{coh}}$}                    & 54.89 & 13.60\\
        \end{tabular}
    \end{ruledtabular}
    \end{center}
    \label{tab:PBdemo}
\end{table}

Ideally, after a sufficiently long gate sequence, the qubit would decohere into a maximally mixed state with $\langle \hat{\sigma}_{z} \rangle = 0$, suggesting no offset ($B = 0$) in the exponential decay fitting. However, in practice, persistent non-zero biases complicate the fitting with an offset term, increasing uncertainty in non-linear least squares fitting and underscoring the necessity to exclude this fixed offset for less fitting errors.

Assuming the qubit reaches its convergence limit after an extensive gate sequence, the survival probability $P$, when targeting the $\left|\text{-}z\right\rangle$ state, should ideally match that when targeting the orthogonal $\left|\text{+}z\right\rangle$ state. This leads to a fitting equation that simplifies the deduction of the offset value $B$. By introducing $\hat{a}$ as the average two measurements:

\begin{align}
    \hat{a} &= \frac{(P_{\text{-}z|\text{-}z}+P_{\text{+}z|\text{+}z})}{2}\notag\\
    &= \frac{(P_{\text{-}z|\text{-}z}-P_{\text{-}z|\text{+}z})}{2}+\frac{1}{2}\label{eq:LNRB},\\
    \overline{a}(m)&= Ap^{m}+\frac{1}{2}\label{eq:LNRBfit},
\end{align}
we eliminate the need to estimate $B$, directly adjusting the offset to 1/2. Here for example, $P_{\text{-}z|\text{+}z}$ represents the probability of measuring state $\text{+}z$ when the overall gate sequence is targeting state $\text{-}z$. To measure $P_{\text{-}z|\text{+}z}$ practically, we replicate the gate sequence, originally designed for $P_{\text{-}z|\text{-}z}$ measurements, with an additional X rotation compiled in the inverse gate standard measurement, enabling probability assessments in both cases.

Reflecting improvements suggested in the literature, we introduce a novel quantity, $\hat{b}$, to model unitarity decay without the offset. This measure is derived from variances in $\hat{\sigma}_k$ measurements, fitted to a simplified decay model, which inherently compensates for and eliminates the offset, thus refining the accuracy of the error estimations:
\begin{widetext}
    \begin{equation}
        \hat{b}(m_i+1) = \frac{1}{n} \sum \limits_{j=1,...,n} \sum \limits_{k=x,y,z} (\langle\sigma_k\rangle_{\mathcal{S}^{m_i+1}_j}^2 - \frac{1}{n}\sum \limits_{j=1,...,n}\langle\sigma_k\rangle_{\mathcal{S}^{m_i+1}_j}^2).  
    \end{equation}
\end{widetext}
Here $\mathbf{S}^{m_i+1}_j$ represents the gate sequence from sub-sequence $G_{m_i}$ within $j$'th PB sequence $\mathcal{S}_j$. This formula calculates the variance in the measurements of $\hat{\sigma}_k$, fitting $\hat{b}$ to a simple decay model:
\begin{equation}
\overline{b}(m) = Au^m.
\end{equation}

Implementing these adjustments in our experimental protocol has proven beneficial, enhancing the precision and reliability of our measurements. By eliminating offset terms from all fitting models, we ensure that our analysis remains robust against the random fluctuations and outstanding fitting uncertainties.

To justify the efficacy of the proposed modifications, we present a proof-of-concept PB experiment. The impact of these improvements on fitting errors is quantitatively demonstrated in Fig. \ref{fig:appx} and Table \ref{tab:PBdemo}. Notably, the coefficient of variation (CV) for each error type is reduced to at least one quarter of its original value, highlighting significant improvements.

\section*{Supplementary Material: Corroborative Experiments}
This section presents three additional experiments that serve as preliminary studies complementing those discussed in the main text. These results further support the conclusions drawn in the main text, albeit with subtlety.

\subsection{Extended Frequency Analysis}
\label{sec:SM1}
\begin{figure}
    \centering
    \includegraphics{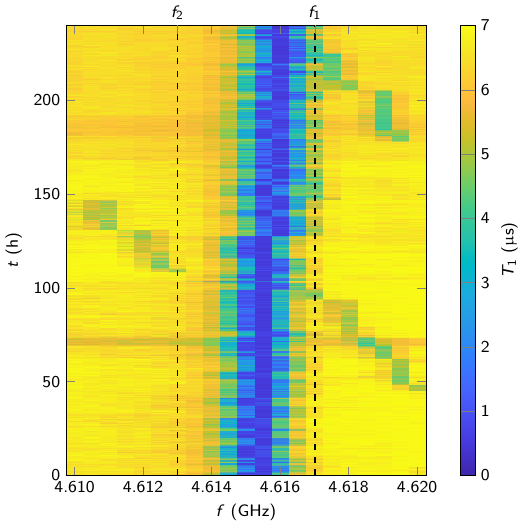}
    \caption{Spectrotemporal chart of $T_1$ vs. $f$ and $t$.}
    \label{fig:2133T1}
\end{figure}

\begin{figure}[b]
    \centering
    \includegraphics{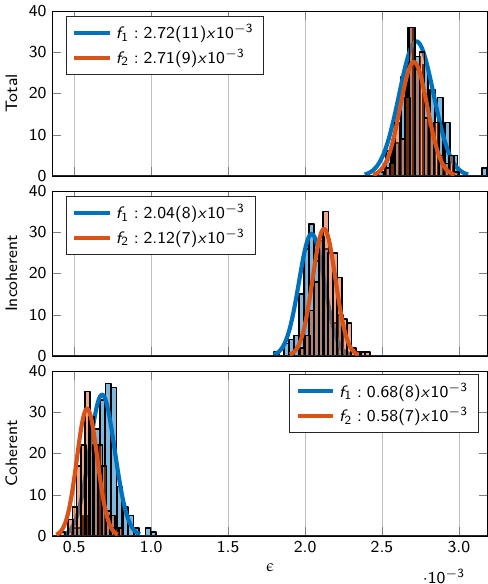}
    \caption{Histograms for time series data.}
    \label{fig:2133hist}
\end{figure}

The first experiment presented here uses exactly the same settings as the two experiments in the main text, but adjusts the PB frequencies, $f_1$ and $f_2$. Similarly as shown in Fig. \ref{fig:2133T1}, $T_1$ fluctuations over time indicate the presence of a strongly coupled TLS centered around $4.615\text{ GHz}$. For this experiment, $f_1$ and $f_2$ are set approximately equidistant from the TLS frequency at around $f_1=4.617\text{ GHz}$ and around $f_2=4.613\text{ GHz}$. Despite a  the TLS center frequency (the $T_1$ dips) slow drifts toward $f_1$.

\begin{figure*}[ht!]
    \centering
    \includegraphics{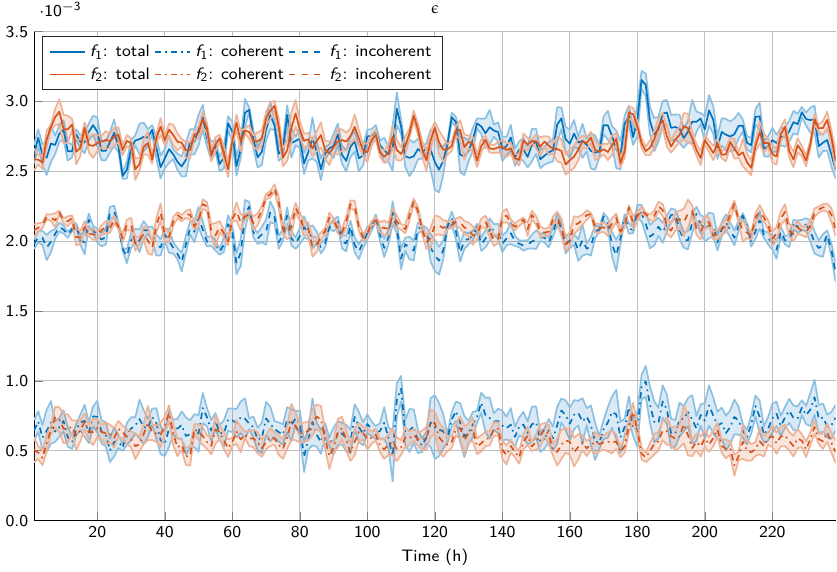}
    \caption{PB time series for different errors in the experiment from Subsection \ref{sec:SM1}, shown with shaded error bars. Line styles and colors are consistent with those used in the main text.}
    \label{fig:2133PB}
\end{figure*}

Figure \ref{fig:2133PB} presents the time series data of this experiment, complemented by histograms Fig. \ref{fig:2133hist} and Table \ref{tab:2133hist}. The data indicates that total errors at both frequencies are still predominantly incoherent. With PB frequencies set equidistant from the TLS, the coherent and total errors at each frequency are nearly identical, aligning with findings from the main text. Additionally, the slight predominance of coherent errors at $f_1$ over $f_2$ correlates with the $T_1$ screening showing a TLS frequency drift towards $f_1$, providing further corroboration with the main text’s conclusions.

\subsection{Experiments with Accelerated Cycle Execution Time}
\label{sec:SM2}
In this section, we discuss the results from two experiments that were conducted with faster cycle executions by omitting concurrent $T_1$ measurements. This modification allows for the accumulation of more measurement cycles within the same time window, enhancing the precision of our fittings and enabling us to capture more rapid error dynamics.

\begin{table*}[ht]
    \caption{Summary statistics of the times series data in Fig. \ref{fig:2133hist}. All error metrics are expressed in the scale of $10^{-3}$.}
    \begin{center}
    \begin{ruledtabular}
        \begin{tabular}{lcccccc}
            \raisebox{0mm}[2.5mm][2.0mm]{} & \multicolumn{2}{c}{$\epsilon$}    & \multicolumn{2}{c}{$\epsilon_\text{inc}$} & \multicolumn{2}{c}{$\epsilon_\text{coh}$} \\
            \raisebox{0mm}[3mm][2mm]{} & $f_1$ & $f_2$ & $f_1$ & $f_2$ &$f_1$ & $f_2$ \\
            \hline
            \hline
            \raisebox{0mm}[3mm][1mm]{Mean $\pm$ SD}  & $2.72\pm0.11$ & $2.71\pm0.09$ & $2.04\pm0.08$ & $2.12\pm0.07$ & $0.68\pm0.08$ & $0.58\pm0.07$\\
            \hline
            \raisebox{0mm}[3mm][1mm]{Median} & 2.71 & 2.70 & 2.04 & 2.11 & 0.68 & 0.58\\
            \hline
            \raisebox{0mm}[3.5mm][0mm]{$25^{th}$ to $75^{th}$ percentile} & 0.04 & 0.00 & 0.06 & 0.01 & 0.01 & 0.05\\
        \end{tabular}
    \end{ruledtabular}
    \end{center}
    \label{tab:2133hist}
\end{table*}

These two experiments are performed consecutively using a setup similar to those in the main text but with PB frequencies set in a different spectra range. Without concurrent $T_1$ measurements, it becomes challenging to directly confirm the presence and central frequency of any strongly coupled TLS. Nevertheless, there are compelling evidences suggesting the transient existence of at least one strongly coupled TLS and a significant telegraphic shift in its frequency during these two experiments.

Figure \ref{fig:2103} and \ref{fig:2111} present the time series data from each experiment, respectively. The data highlight several noteworthy patterns: 
\begin{enumerate}
    \item Incoherent errors still predominantly contribute to the total errors throughout most of the time, consistent with our other experiments. 
    \item Coherent errors are particularly sensitive to telegraphic noise, exhibiting more variability compared to the relatively stable incoherent errors. In fact, almost all time fluctuations in total errors originate from coherent errors. This observation is also consistent with findings from other experiments.
    \item Despite running consecutively, a full recalibration--adjusting drive frequencies, drive pulse amplitudes and DRAG parameter $\lambda$--is conducted just before starting the second experiment. This recalibration significantly reduced the initial coherent errors at the beginning of the second experiment, while the incoherent errors remained at levels similar to those observed at the conclusion of the first experiment. This demonstrating the effectiveness of recalibration in reducing coherent errors (control error) and underscores the robustness of the error classification and reliability of the PB protocol.

\begin{figure*}
    \centering
    \includegraphics{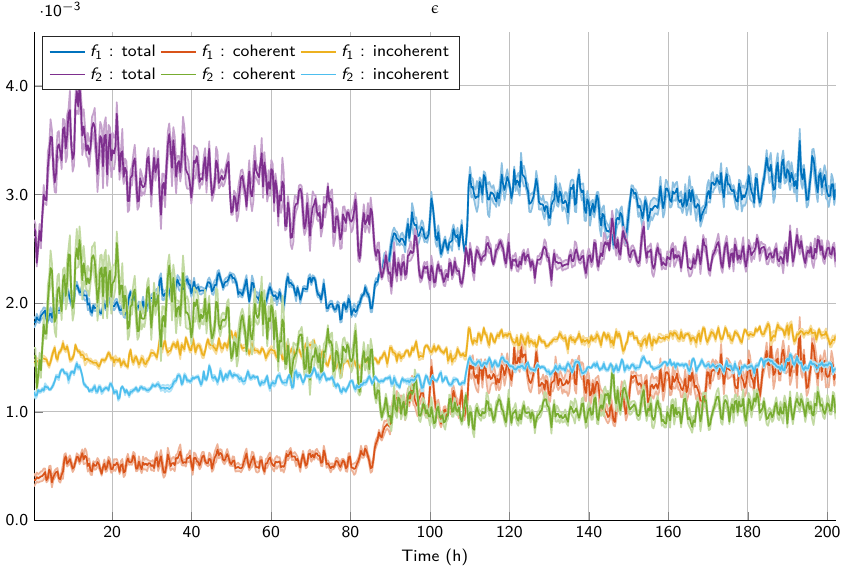}
    \caption{PB time series for different errors in the first experiment from Subsection \ref{sec:SM2}, shown with shaded error bars. Different line styles and colors are used to enhance the graphic presentation. Error bars indicate 68\% confidence intervals ($1\sigma$).}
    \label{fig:2103}
\end{figure*}

\begin{figure*}
    \centering
    \includegraphics{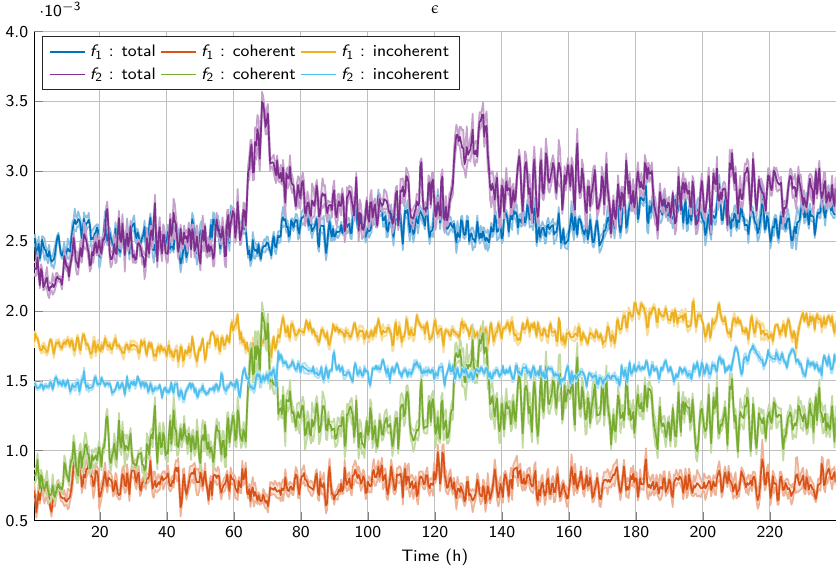}
    \caption{PB time series for different errors in the second experiment from Subsection \ref{sec:SM2}, shown with shaded error bars.}
    \label{fig:2111}
\end{figure*}

\end{enumerate}

\end{document}